\documentclass[%
 twocolumn, 
 amsmath,amssymb,
floatfix,
]{revtex4-1}

\usepackage{graphicx}
\usepackage{dcolumn}
\usepackage{hyperref}
\usepackage{pgfplots}
\usepackage{tikz}
\usetikzlibrary{decorations.markings}
\usepackage{import}
\usepackage[version=4]{mhchem}

\usepackage{braket}

\usepackage{wasysym}

\pgfplotsset{compat=1.17}

\usepackage{xcolor}

\newcommand{\llangle}{\langle \langle}
\newcommand{\rrangle}{\rangle \rangle}

\begin{document}


\title{Vison crystal in quantum spin ice on the breathing pyrochlore lattice}

\author{Alaric Sanders}
\email{als217@cam.ac.uk}
\affiliation{
TCM Group, Cavendish Laboratory, University of Cambridge, Cambridge CB3 0HE, United Kingdom
}
\author{Claudio Castelnovo}
\affiliation{
TCM Group, Cavendish Laboratory, University of Cambridge, Cambridge CB3 0HE, United Kingdom
}

\date{\today}

\begin{abstract}
Recent excitement in the quantum spin ice community has come from the experimental discovery of pseudospin-$\frac{1}{2}$ breathing pyrochlores, including Ba$_3$Yb$_2$Zn$_5$O$_{11}$, in which inversion symmetry is broken by the `up' and `down' tetrahedra taking different physical sizes. We show here that the often-neglected $J_{z\pm}$ coupling between Kramers ions, in combination with the breathing nature of the lattice, can produce an imaginary ring flip term.
This can lead to an unconventional `$U(1)_{\pi/2}$ phase', corresponding to a maximally dense packing of visons on the lattice. Coherent dynamics persists in all phases, together with its emergent QED description, in a manner reminiscent of fragmentation in spinon crystals. We characterize the enlarged QSI phase diagram and its excitations, showing that the imaginary ring flip acts both as a chemical potential for visons and as an effective three-photon vertex akin to strong light-matter coupling. 
The novel coupling causes a structured high-energy continuum to emerge above the photon dispersion, which is naturally interpreted as three photon up-conversion in a nonlinear optical crystal.
\end{abstract}

\maketitle

%
%

\section{Introduction}
\label{sec:intro}
Quantum spin ice (QSI) is one of the best known three dimensional models realising quantum spin liquid phases. In addition to the gapped spinon excitations of classical spin ice, QSI is characterized by the presence of an emergent $U(1)$ gauge field bearing a striking resemblance to compact scalar QED~\cite{hermelePyrochlorePhotonsSpin2004}, complete with emergent analogues of Dirac monopoles (visons) and gapless gauge photons. Here we adopt the `physical' naming convention in which the spinons are said to carry magnetic charge in the emergent QED, while visons carry electric charge. This convention is adopted to reflect the well-established (real) magnetic charge of the spinons in spin ice materials~\cite{castelnovoMagneticMonopolesSpin2008,castelnovoDebyeHuckelTheorySpin2011}, and the more recent suggestion that visons may experience Lorentz force in the presence of an externally applied magnetic field~\cite{laumannHybridDyonsInverted2023, zhangTopologicalThermalHall2020}.
Correspondingly, we shall use the language of Refs.~\onlinecite{castelnovoMagneticMonopolesSpin2008,
kwasigrochSemiclassicalApproachQuantum2017}, in
which the $S^z$ operators correspond to the magnetic field, while canonically conjugate rotor variables $a$, $[S^z, a]=-i$, are understood as the $U(1)$ vector potential of an emergent 
electric 
field~\footnote{Note that there exists an equally valid alternative convention in which $S^z$ is called the electric field~\cite{hermelePyrochlorePhotonsSpin2004,paceDynamicalAxionsQuantum2021}.}.

Recent progress~\cite{savaryQuantumSpinIce2016} has been made on the generalization of quantum spin ice models to the breathing pyrochlore lattice, motivated on the experimental side by breathing QSI candidate materials such as \ce{Ba_3Yb_2Zn_5O_11}~\cite{rauAnisotropicExchangeDecoupled2016,hakuLowenergyExcitationsGroundstate2016, heNeutronScatteringStudies2021,
okamotoBreathingPyrochloreLattice2013,ghoshBreathingChromiumSpinels2019,okamotoBreathingPyrochloreLattice2013,
okamotoMagneticStructuralProperties2018}, and on the theoretical side by intriguing predictions of
axion-like excitations~\cite{paceDynamicalAxionsQuantum2021} and higher-rank $U(1)$ gauge theories~\cite{yanRankSpinLiquid2020,bulmashHiggsMechanismHigherrank2018}. 

In this work, we study a realistic model of breathing quantum spin ice, showing that within the spin liquid phase the breathing anisotropy and generic nearest-neighbor couplings can conspire to create an all-in, all-out (AIAO) bias in the electric field. If tuned beyond a critical strength in parameter space, we demonstrate that such a bias field induces a phase transition and the emergence of a vison crystal, or equivalently, a $U(1)_{\pi/2}$ phase. 

The AIAO bias echoes an existing model in the magnetic sector, which we summarize here as a useful analogy for our generalization to the electric sector. In certain rare-earth pyrochlore iridates (\ce{RE_2Ir_2O_7}, RE$\,\in\,\{\text{Ho}, \text{Dy}, ...\}$), classical spin ice is interpenetrated by a pyrochlore lattice of AIAO-ordered iridium atoms that generate a staggered, AIAO magnetic field $h$ aligned with the local $z$ axis on every pyrochlore site $j$~\cite{pearceMonopoleDensityAntiferromagnetic2022,
lefrancoisAnisotropyTunedMagneticOrder2015, lefrancoisFragmentationSpinIce2017,
antonovPyrochloreIridatesElectronic2020, sagayamaDeterminationLongrangeAllinallout2013,
guoDirectDeterminationSpin2016, guoDirectDeterminationSpin2016, lefrancoisFragmentationSpinIce2017},
\begin{equation}
\label{eq:shifted_pyro}
H_{\ce{RE_2Ir_2O_7}} = \sum_{t \in \{\text{Tetrahedra} \} } \frac{J_{zz}}{2} \left[ \sum_{j \in t} \left( S^z_j - \eta_t \frac{h}{2J_{zz}} \right)
\right]^2 
\, ,
\nonumber
\end{equation}
where $\eta_t$ is +1 on the $A$ tetrahedral sublattice and -1 on the $B$ sublattice. Both inversion and time reversal symmetry are explicitly broken, prejudicing `in' ($-$) and `out' ($+$) spinons to lie on different sublattices. The staggered field $h$ drives a strongly first order monopole condensation transition~\cite{chenMagneticMonopoleCondensation2016} from the divergence free 2-in, 2-out (2I2O) phase to a 3-in, 1-out (3I1O) spin-fragmented phase, ultimately reaching an AIAO-ordered phase if $h$ is further increased~\cite{cathelinFragmentedMonopoleCrystal2020,lefrancoisSpinDecouplingStaggered2019, lefrancoisFragmentationSpinIce2017, petitObservationMagneticFragmentation2016}.

We show that an analogue of the AIAO magnetic bias field is generally present in the electric sector of a breathing spin-$\frac{1}{2}$ quantum spin ice. The role of the iridium AIAO magnetic field is played, subtly, by the $J_{z\pm}S^z_iS^{\pm}_j$ term (acting within conserved spinon subspaces), assigning different phases to spinon hopping processes $S^{\pm}_j$ depending on the state $S^z_i$ of the spin at site $i$. When these terms are used to transport a virtual spinon around a closed loop in the standard fashion of a ring flip process, a nonvanishing phase correction emerges in the effective Hamiltonian.
The main consequences of this novel contribution to the Hamiltonian may be summarized as follows:
\begin{enumerate}
    \item Nearest-neighbor microscopic spin interactions on the Kramers breathing pyrochlore generically generate an AIAO electric bias field.
    \item Numerics reveal that this electric bias field selects between four possible phases: the known $U(1)_0$ and $U(1)_\pi$ phases, and two unconventional, fragmented vison crystal states which we refer to as the $U(1)_{\pm\pi/2}$ phases.
    \item These phases are separated by lines of liquid-gas-like first order phase transitions, terminating at a critical end point and becoming all continuously connected via the disordered high-temperature phase.
    \item The bias field is decoupled from the emergent photon dynamics to quadratic order. 
    \item The leading order correction to the photon propagator comes at cubic order in the emergent gauge field $a$, representing a three-photon vertex correction.
    \item Visons interact via an emergent Coulomb law, with the bias field acting (within a large-$S$ description) only as a chemical potential for visons, prejudicing them to lay on a certain sublattice set by the original breathing anisotropy.
\end{enumerate}

The paper is structured as follows. We introduce the model in Sec.~\ref{sec:model}. The microscopic perturbation theory responsible for the AIAO offset is outlined in Sec.~\ref{sec:perturb}. 
Sec.~\ref{sec:staggered_intro} presents certain symmetries of the minimal model, which we employ to explore the parameter space more efficiently. We then study in Sec.~\ref{sec:phasedia} the phase diagram of this AIAO-biased quantum spin ice using semiclassical simulations~\cite{kwasigrochSemiclassicalApproachQuantum2017,szaboSeeingLightVison2019}.
We find a sharp, liquid-gas like phase transition between the known $U(1)_0$ QSI phase and an unconventional $\pi/2$-flux phase. This transition appears to terminate at a critical endpoint, which we investigate using finite-size scaling.
Sec.~\ref{sec:uniform} presents static and dynamic structure factors within the $U(1)_{\pi/2}$ phase. In Sec.~\ref{sec:vison_interactions} we obtain a functional form fo the vison-vison interaction potential in the new regime.
Finally, we draw our conclusions in Sec.~\ref{sec:conclusions}. 
%
%

\section{Breathing pyrochlore quantum spin ice}
\label{sec:model}
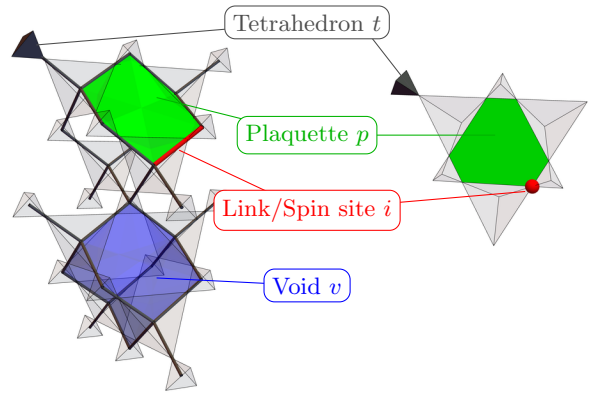
\begin{figure}
    \begin{tikzpicture}
    \node[anchor=south west,inner sep=0] at (0,0){\includegraphics{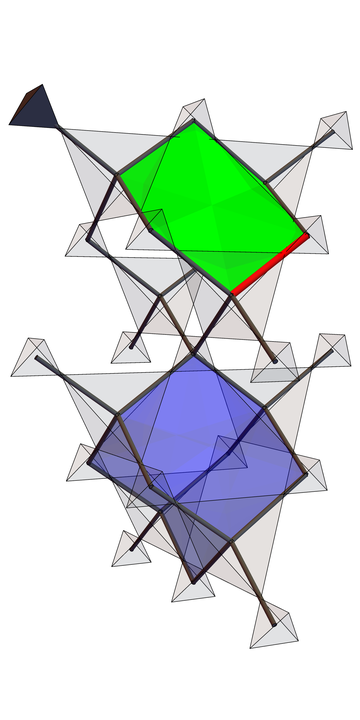}};
    \node[anchor=south west,inner sep=0] at (5,2.2){\includegraphics{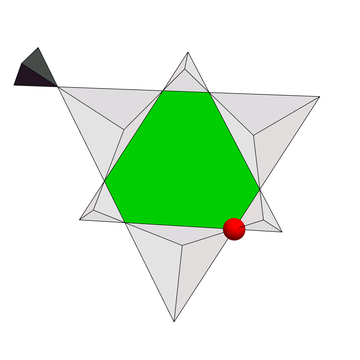}};
    \draw[black!70!white]   (0.4,5.2)--(4,5.5)node[fill=white, draw,rounded corners]{Tetrahedron $t$} --++(0.7,0)-- (5.4,4.6);
    \draw[green!70!black] (2,4.5)--(4,4) node[fill=white,draw,rounded corners] {Plaquette $p$} -- (6.5,4);
    \draw[red] (2.3,3.8) -- (4,3) node[fill=white,draw,rounded corners] {Link/Spin site $i$} -- (6.9,3.25);
    \draw[blue] (2,2.1) -- (4 ,2)node[fill=white,rounded corners,draw] {Void $v$};
    \end{tikzpicture}
    \caption{Illustration of the pyrochlore and diamond lattices. Gray tetrahedra share corners which form the pyrochlore sites $i$ (red sphere, right), which are in one-to-one correspondence with the links (red line, left) of the diamond lattice. Each diamond plaquette $p$ (green even-sides hexagon, left) is similarly in unique correspondence with the hexagonal plaquettes of the pyrochlore lattice (green alternating-sided hexagon, right). Four pairwise adjacent plaquettes enclose a void $v$ (purple region, bottom left).}
    \label{fig:lattice_objects}
\end{figure}

%
%
%
We begin by choosing a notation with which to index the sites of the breathing pyrochlore lattice. The spins of the model, residing on the pyrochlore sites, can be conveniently thought of as living on the links of a parent diamond lattice, as illustrated in Fig.~\ref{fig:lattice_objects}
~\footnote{The original pyrochlore lattice is medial -- or bond-dual -- to the diamond lattice.}. 
The pyrochlore tetrahedra are centered on the diamond sites, and the smallest closed loops of links form (buckled) hexagonal plaquettes. Four such plaquettes can be chosen to enclose three-dimensional volumes, which we will refer to as voids~\footnote{The voids are also known as `dual tetrahedra'~\cite{hermelePyrochlorePhotonsSpin2004} or `cells'~\cite{weberGaugeTheorySpatial2013}.}. 
It will be most convenient to use the language of pyrochlore sites and tetrahedra for the perturbative derivation of the effective Hamiltonian in Sec.~\ref{sec:perturb}, while the effective theory of the subsequent sections is more elegantly described in the diamond lattice language.

We obtain the Hamiltonian of breathing quantum spin ice
by modifying the standard interaction terms in non-breathing, Kramers pyrochlore spin ice ($J_{zz}$, $J_{\pm}$, $J_{z\pm}$, $J_{\pm\pm}$) to take different strengths on the two inequivalent tetrahedra, which we name $A$ and $B$~\cite{savaryQuantumSpinIce2016,hakuCrystalFieldExcitations2016,huangQuantumSpinIces2014,onodaQuantumFluctuationsEffective2011,rossQuantumExcitationsQuantum2011,onodaEffectiveQuantumPseudospin12011,onodaQuantumMeltingSpin2010,curnoeStructuralDistortionSpin2008}:
\begin{align}
\label{eq:Hamiltonian_full}
H &= \sum_{\sigma \in \{A,B\}} H_{zz}^\sigma + H_\pm^\sigma + H_{z\pm}^\sigma + H_{\pm\pm}^\sigma
\\
H_{zz}^{\sigma} &= \sum_{\langle ij\rangle_\sigma}  J_{zz,\sigma} S^z_i S^z_j
\nonumber \\
H_{\pm}^{\sigma} &= \sum_{\langle ij\rangle_\sigma}-J_{\pm,\sigma} \left( S^+_iS^-_j + S^-_iS^+_j \right) 
\nonumber \\
H_{z\pm}^{\sigma} &= \sum_i  \sum_{ \langle j\to i \rangle_\sigma } J_{z\pm,\sigma}\zeta_{ij} S^z_j S^+_i + h.c. 
\nonumber \\
H_{\pm\pm}^{\sigma} &= \sum_{\langle ij\rangle_\sigma} J_{\pm\pm,\sigma} \left( \gamma_{ij} S^+_i S^+_j + \gamma_{ij}^*S^-S^- \right) 
\, , 
\nonumber
\end{align}
where, if $\mu(i)=1,2,3,4$ denotes the pyrochlore sublattice of link $i$, then 
\begin{align}
\gamma_{\mu(i)\mu(j)}^* &= -\zeta_{\mu(i)\mu(j)} 
\nonumber \\
&=
\left(\begin{array}{cccc}
  0 & 1 & e^{2\pi i/3} & e^{-2\pi i/3} \\
  1 & 0 & e^{-2\pi i/3} & e^{2\pi i/3} \\
  e^{2\pi i/3} & e^{-2\pi i/3} & 0 & 1 \\
e^{-2\pi i/3} & e^{2\pi i/3} & 1 & 0
\end{array}\right)
\, , 
\end{align}
$\sigma$ is a dummy index for the diamond (i.e., tetrahedral) sublattices $A,B$ and $\langle ij \rangle_\sigma$ denotes a pyrochlore bond belonging to a tetrahedron on sublattice $\sigma$. 
The sum over $i$ in the fourth line runs over all pyrochlore sites, with the subsequent sum $\langle j \to i\rangle_\sigma$ running over the six (three $A$, three $B$) nearest neighbors of $i$.
Note that each spin component is understood with respect to the local (pyrochlore sublattice dependent) axes tabulated in Appendix~\ref{app:geometry}. Our attention is restricted to the case of Kramers doublets; the additional conditions satisfied by non-Kramers and dipolar-octupolar doublets are given in Table~\ref{tab:symmetry_cases} for completeness.

We suppose that both $J_{zz,A}$ and $J_{zz,B}$ are much stronger than the remaining couplings. Without loss of generality let the $A$-sublattice coupling be stronger, $J_{zz,A} \gtrsim J_{zz,B} > 0$. If we neglect all other terms, the Hamiltonian becomes
\begin{equation}
\label{eq:disorder_csi}
H_{0} = \sum_{t \in \{ \text{Tetrahedra} \} } 
J_{zz,\sigma(t)}
\left( \sum_{i \in t} S^z_i \right)^2
\, ,
\end{equation}
and the ground states belong to the canonical 2I2O ice ensemble. The energy of a spinon excitation becomes sublattice dependent, rendering the $B$-sublattice spinons (irrespective of their charge) less energetically expensive. 
The spinon gap is therefore controlled by $J_{zz}^B$. 
In this respect, we note that the known breathing spin-$\frac{1}{2}$ pyrochlore material to date, \ce{Ba_3Yb_2Zn_5O_11}, suffers from an exceedingly small value of $J_{zz,B}$, leaving the $A$ tetrahedra paramagnetically decoupled down to $\sim 0.1$~K~\cite{dissanayakeUnderstandingMagneticProperties2022}.
%
%

\section{Perturbation Theory}
\label{sec:perturb}
We next revisit the perturbation theory of Hermele, Fisher and Balents~\cite{hermelePyrochlorePhotonsSpin2004}, starting from the general symmetry-allowed nearest-neighbor Hamiltonian in Eq.~\eqref{eq:Hamiltonian_full} compatible with the effective spin-$1/2$ $C_{3v}$ symmetry. 

In the standard fashion, we consider $J_{zz,A} \gtrsim J_{zz,B} \gg |J_{\pm,\sigma}| \sim |J_{z\pm,\sigma}| \sim |J_{\pm\pm}|$.
We construct an effective Hamiltonian that operates solely within the 2I2O spin ice ensemble using the standard expansion
\begin{align}
    H_{\text{eff}} = P V \sum_{k=0}^\infty \left[ \frac{1-P}{-H_0} V \right]^k P 
\, ,  
\label{eq:eff_ham_expansion}
\end{align}
where $P$ projects onto the 2I2O ice manifold, and the interaction term is $V = H_\pm^A + H_\pm^B + H_{z\pm}^A + H_{z\pm}^B$. 
We also remind in passing that second order perturbation theory is known to only generate farther range Ising (namely, $S^z$) interactions that -- when sufficiently small -- do not disrupt the QSL phase outright, but merely renormalize the emergent speed of light~\cite{paceEmergentFineStructure2021, savaryCoulombicQuantumLiquids2012}.
In what follows, we suppose that the system remains continuously connected to the QSL phase, with the Ising interactions simply weighting the state distribution.
We similarly neglect any four-$S^z$ and longer range Ising terms (e.g., those dubbed $\gamma$ and $I_2$ in Table~\ref{tab:terms}) that occur at higher orders.

\begin{figure}
	\includegraphics{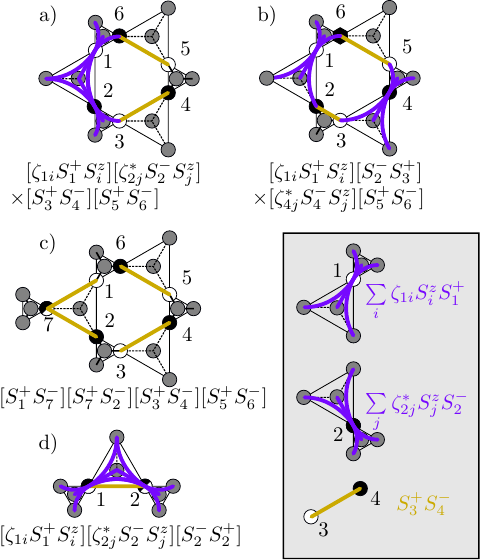}
	\caption{Selection of processes (i.e., ring flip loops and retracing paths) appearing up to fourth order (see also Table~\ref{tab:terms}), together with their associated (unordered) operator sets. Black and white symbols denote spins that are flipped in the process, respectively up-to-down and down-to-up. Circles denote pyrochlore sites, i.e. spins. Spins that are flipped twice (returning to their original state) are half-filled in black and white. Gray spins are not flipped by the process. Curved purple lines indicate the six nearest neighbors of a ladder operator $S^\pm$, over which the sums over $i$,$j$ run.
    Loops a) and b) illustrate the two classes of terms arising from $H_{\pm}^2 H_{z\pm}^2$ that are responsible for generating complex ring flips, as explained in the main text. Loop c) is a fourth-order `teardrop' correction to the conventional ring flip term. Self-retracing loop d) generates only an Ising (namely, $S^z$) interaction. Note that the smaller (i.e., $A$) tetrahedra are taken to face out of the plane of the page, as in right panel of Fig.~\ref{fig:lattice_objects}.}
\label{fig:terms}
\end{figure}

The processes generated at third and fourth order have sufficiently many vertices to create a virtual spinon pair, transport the spinons around a non-trivial (i.e., non-self-retracing) loop on the diamond lattice, and annihilate them. The prototypical example of such a process is 
the well-known ring flip term from $H_\pm^3$, which generates $g^{(3)} = \frac{3J_{\pm, A}^3}{4 J_{zz, B}^2} + \frac{3J_{\pm, B}^3}{4 J_{zz, A}^2}$ and the corresponding $\mathcal{O}_{\hexagon} = S_1^+ S_2^- S_3^+ S_4^- S_5^+ S_6^-$ operator (where $1,2,\ldots,6$ label the six sites around the hexagon $\hexagon$). Another example at higher order comes from $H_{\pm}^4$, which generates `teardrop' ring flips (see Fig.~\ref{fig:terms}d), as well as some additional trivial energy shifts. 


An illustrative selection of the resulting terms is given in Fig.~\ref{fig:terms}, and detailed in Table~\ref{tab:terms}. 
The coupling $H_{\pm\pm}$ can be shown (see Appendix~\ref{app:details}) to generate only multi-$S^z$ interactions up to fifth order, and we therefore neglect it for the purpose of our discussion.
\begin{table}[h!]
    \begin{tabular}{l|l} 
        Origin & Nontrivial Terms \\  
        \hline 
		$H_{z\pm}^2$ & $ I_2^{(2)} \sum_{\llangle ij \rrangle} S^z_i S^z_j$ \\
		$H_{\pm}^3$ & $ g^{(3)}\sum_{p}(\mathcal{O}_{p} +
        \mathcal{O}_{p}^\dagger) $\\
		$ H_{z\pm}^2 H_\pm$ & $I_2^{(3)} \sum_{\llangle ij \rrangle} S^z_iS^z_j$\\
  	$ H_{z\pm}^2 H_\pm$ & $ \gamma^{(3)}
		\sum_{\langle i j k l\rangle} S^z_i S^z_j S^z_k S^z_l$ \\
		$ H_{\pm}^4$ & $g^{(4)}  \sum_{p}  \mathcal{O}_{p} + \mathcal{O}^\dagger_{p} $ \\
		$ H_{z\pm}^2 H_{\pm}^2$ & $ ig'^{(4)}\sum_{p} \mathcal{O}_{p} -
	\mathcal{O}^\dagger_{p}  $
    \end{tabular}
\caption{Selected terms generated to fourth order in perturbation theory on the breathing pyrochlore lattice, including second-neighbor Ising interactions $\llangle ij \rrangle$ and four spin interactions between nearest-neighbour self-avoiding lines $\langle ijkl \rangle$. Expressions for all corresponding effective coupling constants in terms of the original interactions are given in Appendix~\ref{app:details}.}
\label{tab:terms}
\end{table}
\begin{figure}
    \includegraphics{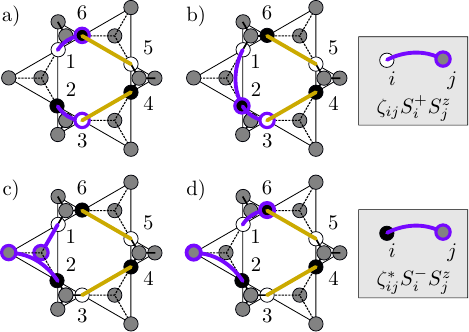}
    \caption{Four of the individual terms contributing to Fig.~\ref{fig:terms}a. The curved purple lines denote a factor of $\zeta_{ij}S^+_iS^z_j$ or $\zeta_{ij}^*S^-_iS^z_j$. As indicated in the main text, a), b) lead to complex ring flips whereas c) leads to real ring flips. The term d) cancels with another term in the expansion and does not contribute.}
    \label{fig:cplx_ring flips}
\end{figure}

Importantly, in addition to the canonical inversion-even $g$ terms, we obtain an imaginary ring flip $i g'\left(\mathcal{O} - \mathcal{O}^\dagger\right)$ from the eight-spin ring exchange $H_{\pm}^2 H_{z\pm}^2$. This term arises from the sum over all strings of the form  $\zeta_{i1}S^z_{i} S^+_1 \zeta^*_{j2} S^z_{j} S^-_2 S^+_3 S^-_4 S^+_5  S^-_6$, where 1-6 refer to the pyrochlore site labeling of Figs.~\ref{fig:terms} and~\ref{fig:cplx_ring flips}. When both $i$ and $j$ lie on the ring (i.e. $i,j \in\{1,2,...,6\}$), as they do in Fig.~\ref{fig:cplx_ring flips}a,b, repeated applications of the identity $S^zS^\pm \equiv \pm S^\pm / 2$ replace both factors of $S^z$ by scalars. The effective operator is then proportional to $\mathcal{O}_{\hexagon}$, with a possibly complex coefficient. Such complex ring flips are generally forbidden by inversion symmetry, but are allowed on the breathing pyrochlore where this symmetry is explicitly broken.

We illustrate here an example of a non-reciprocal process responsible for these imaginary ring flips. Consider the process shown in Fig.~\ref{fig:cplx_ring flips}a, which represents a sum of all 4! orderings of the operator set $\{\zeta_{16} S^+_1 S^z_{6},\ \zeta^*_{23}  S^-_2 S^z_{3},\ S^+_3 S^-_4,\ S^+_5  S^-_6\}$. In exactly half of these, the $S^-_6$ is to the left of the term containing $S^z_6$, contributing a factor of $-S^z_6 / 2$. The other 12 terms have the opposite sign. Similarly, half of the terms in the sum have $S^z_3$ to the left of $S^+_3$, giving an additional permutation-dependent sign. It can be checked explicitly that the four cases,
\begin{align*}
\begin{array}{l|cc}
& S^z_3S^+_3 & S^+_3 S^z_3\\
\hline
S^z_6S^+_6 & + & - \\
S^+_6S^z_6 & - & + \\
\end{array}
\end{align*}
are realized by six terms each. As all other terms in the string commute, this would appear to suggest that the overall sum vanishes exactly, and that processes like those of Fig.~\ref{fig:cplx_ring flips} do not contribute to the ring flip. This intuition has, however, failed to account for the factors of $H_0$ in the denominator of the expansion in Eq.~\eqref{eq:eff_ham_expansion}. Reintroducing the propagators $\frac{1}{-H_{0}}$, where $H_0$ is the Ising Hamiltonian~\eqref{eq:disorder_csi}, gives 
\begin{align}
    S_4^-S_3^+ \frac{1}{-H_0}\zeta_{23}^* S_2^- S_3^z 
    \frac{1}{-H_0}
    \zeta_{16} S_1^+S_6^z \frac{1}{-H_0} S_6^-S_5^+ &\label{eq:m1}
    \\
    + S_4^-S_3^+\frac{1}{-H_0}\zeta_{23}^* S_2^- S_3^z
    \frac{1}{-H_0} 
    S_6^-S_5^+ \frac{1}{-H_0} \zeta_{16} S_1^+S_6^z & \
    \, . 
    \label{eq:m2}
\end{align}
In the term~\eqref{eq:m1}, the rightmost factor of $H_0$ measures two spinons, one on each of the $A$ and $B$ tetrahedral sublattices, giving a factor of $2/(J_{zz}^A + J_{zz}^B)$. The corresponding propagator in~\eqref{eq:m2} instead measures two spinons on the $B$ sublattice, giving a factor of $1/J_{zz}^B$. The terms therefore do not cancel. 

There is an important subtlety associated with orienting the ring flip operators $\mathcal{O}$ on the breathing pyrochlore lattice. The loops in Fig.~\ref{fig:terms} are oriented with the smaller $A$ tetrahedra facing out of the plane of the page, with ladder operators arranged such that a counterclockwise traversal of the ring raises (lowers) the spins encountered when moving from an $A$ ($B$) tetrahedron to a $B$ ($A$) tetrahedron. The opposite convention would generate instead $\mathcal{O}^\dagger$. For consistency, it follows that $g$ must be even under exchange of tetrahedron sublattices, while $g'$ must be odd.

For completeness, we mention that these $H_{\pm}^2 H_{z\pm}^2$-type ring processes also generate ring flips of the form $S^zS^z\mathcal{O}$ where the dangling $S^z$ factors are not part of the ring ($i,j \, \notin \{1,2,\ldots,6\}$ in the notation above, see Fig.~\ref{fig:cplx_ring flips}c). We interpret these as ring flips modulated by $S^zS^z$ correlations in the local environment, which we elaborate on in Appendix~\ref{app:details}. We will ultimately drop such eight-spin operators, restricting our attention to the `pure' ring flips~\cite{bentonSeeingLightExperimental2012,kwasigrochSemiclassicalApproachQuantum2017,kwasigrochVisongeneratedPhotonMass2020}. Note that for any $C\in\mathbb{C}\setminus\{0\}$, terms of the form $C S^z \mathcal{O} + C^*S^z\mathcal{O}^\dagger$ violate time reversal symmetry and do not appear in the effective Hamiltonian; it follows that 1-on, 1-off complex loops, as the case illustrated in Fig.~\ref{fig:cplx_ring flips}d, give no overall contribution.

If we further parameterize the couplings by introducing the anisotropy parameter $\kappa$, where $(1+\kappa)/(1-\kappa) = J_{zz,A}/J_{zz,B} = J_{z\pm,A}/J_{z\pm, B} = J_{\pm, A}/J_{\pm, B}$, it is possible to give an asymptotic expression for $g,g'$ in the vicinity of $\kappa=0$. Denoting the average couplings with an overline (e.g., $\overline{J}_{zz} = ( J_{zz,A} + J_{zz,B} )/2$), we have
\begin{align}
    g  \sim& -\frac{6}{\overline{J}_{zz}^3}\left[36 \overline{J}_\pm^4 - \frac{31}{4}\overline{J}_\pm^2\overline{J}_{z\pm}^2 + 2 \overline{J}_\pm^3 \overline{J}_{zz} \right]  + O\left(\kappa^2\right)
    \label{eq:g_linearised}
    \\
    g' \sim& -\frac{51\sqrt{3}\ \overline{J}_\pm^2\overline{J}_{z\pm}^2}{2\overline{J}_{zz}^3}  \kappa
    + O\left( \kappa^3 \right)
\, . 
    \label{eq:gp_linearised}
\end{align}
As required by the definition of $\mathcal{O}$, it is true to all orders that $g'$ is an odd function of $\kappa$, whereas $g$ is even.

Take note that the fourth-order contribution to $g$ has a far larger combinatorial prefactor than the third-order term (216 as opposed to 12). This is suggestive of a slowly converging perturbative series in cases where $\overline{J}_\pm/\overline{J}_{zz}$ is moderately large. Indeed, recent gauge mean field theory calculations~\cite{desrochersSpectroscopicSignaturesFractionalization2023} suggest that the $U(1)_\pi$ phase remains stable up to $J_\pm \approx J_{zz}$. 
%
%

\section{The $g+ig'$ model
\label{sec:staggered_intro}}
In the rest of the discussion, we restrict our attention to a minimal model containing $g$ and $g'$ as the sole parameters: 
\begin{equation}
H = -\sum_p (g+ig') \mathcal{O}_p + h.c. 
\, . 
\label{eq:heff}
\end{equation}
The sign of $g'$ reflects a choice of labeling for the $A$ and $B$ tetrahedra. The earlier convention $J_{\pm,A} \ge J_{\pm, B}$ renders, without loss of generality, $g'$ non-negative. 
It will be later convenient to express $g+ig' = \rho_g e^{i \phi_g}$ (recall that $g$ and $g'$ are real). 
%
%

\subsection{Large-$S$ limit and semiclassical theory}
\label{ssec:u1-pi2}
We apply the canonical Villain expansion~\cite{villainInsulatingSpinGlasses1979}, introducing an operator $a$ canonically conjugate to $S^z$ (with $[a, S^z] = i$). As the $a$ variables are quantum rotors, they are compact; our work, following standard convention, understands $a$ to be an operator on the Hilbert space $L^2(U(1))$ of square integrable functions on the periodic $[0,2\pi)$ interval, which will ultimately be represented as a c-number in $U(1) = [0,2\pi)$ in the semiclassical theory. Ladder operators for spin $S$ are then replaced by the expression 
\begin{equation}
\label{eq:villain_expansion}
    S^+ = e^{ia/2} \sqrt{(\tilde s)^2 - (S^z)^2} \, e^{ia/2} 
    \, ,
\end{equation}
where $\tilde s = S + 1/2$  (see also Refs.~\onlinecite{hermelePyrochlorePhotonsSpin2004,bentonSeeingLightExperimental2012,kwasigrochSemiclassicalApproachQuantum2017}). 
We expand $\mathcal{O}_p$ to leading order in $1/\tilde{s}$ to obtain a modified form of the canonical $U(1)$ lattice gauge theory,
\begin{align}
    \frac{H}{\tilde s^6} &= - \sum_p \left[
    \vphantom{\sum}
    g \cos( \nabla \times_p a) + g' \sin(\nabla \times_p a)
    \right] 
\label{eq:large_S_ham} \\
    &= -\sum_p \rho_g \cos( \nabla \times_p a -\phi_g) 
    \, , 
\end{align}
where we introduced the lattice curl $\nabla \times_p a = \sum_{j\in p} (-)^j a_j$, with an identical sign convention to that which was used in the definition of $\mathcal{O}$, 
chosen to respect the orientation of the plaquette $\mathcal{O}_p$ operators. 
In the second line above we reparameterized for convenience the coupling constants as $g = \rho_g\cos(\phi_g)$ and $g' = \rho_g\sin(\phi_g)$. 
With this definition, it becomes clear that $a$ is a compact $U(1)$ gauge field - $\nabla\times a$ is the discrete line integral of $a$ around a closed loop, i.e. a gauge flux.

Notice the subtle importance of working first on the perturbation theory of the spin-1/2 system, to produce an effective Hamiltonian that contains terms with the same number of spin operators, and then taking the large-$S$ limit. While we have combined third and fourth order terms, which nominally have $6$ and $8$ spin operators respectively, we argued in Sec.~\ref{sec:perturb} that the interesting and novel contributions from the latter arise when $2$ of the $8$ spin operators (namely, the $S^z$ terms) have a known expectation value and can therefore be included in the numerical coefficients. This reduces the $8$ spin operators of interest to $6$ spin operators, thence Eq.~\eqref{eq:heff}. Turning then to large-$S$ theory produces a well-defined Hamiltonian where all the contributions are of order $\tilde{s}^6$. Of course, the fact that some terms come from third order perturbation theory, and some from fourth order, remains reflected in the parametric form of the Hamiltonian coefficients. 

The Hamiltonian in Eq.~\eqref{eq:large_S_ham} is cast in terms of a $U(1)$ lattice gauge field $a$. It will be useful to the later discussion to introduce the unitary $\mathcal{C}$, which rotates all spins about the local $x$ axis.
\begin{align}
\mathcal{C} :\ 
\begin{pmatrix}
  S^z_i\\S^\pm_j
\end{pmatrix} 
&\quad \mapsto \quad 
\begin{pmatrix}
  -S^z_i\\S^\mp_i
\end{pmatrix}
\, . 
\label{eq:C_definition} 
\end{align}
This operation reverses all $S^z$ moments, inverting the sign of all magnetic monopoles. Further, its action on $S^\pm$, 
\begin{align}
\mathcal{C}[S^+] = S^- \quad \Leftrightarrow \quad \mathcal{C}[e^{ia}] = e^{-ia}
\, , 
\end{align}
suggests that $\mathcal{C} a = -a$, reversing the electric field strength. Inspired by the observation that $\mathcal{C}$ reverses both electric and magnetic field lines, we name this operation charge conjugation.

It follows immediately from this discussion that $\mathcal{C}$ maps $(g, g') \mapsto (g, -g')$. Further, it may be straightforwardly verified that, in the effective Hamiltonian, $\mathcal{C}$ is equivalent to the formal inversion operation in which $A$ and $B$ sublattice labels are interchanged, in a QSI analogue of QED's charge-parity symmetry.

Note that $H_{zz}$ and $H_{\pm}$ are invariant under $\mathcal{C}$. It is therefore a symmetry of all pure-XXZ spin models, and indeed all dipolar-octupolar spin ice Hamiltonians. This highlights breathing Kramers QSI as a promising system for generating complex ring flips
~\footnote{In principle, $J_{\pm\pm}$ in a non-Kramers material also violates charge conjugation, but such terms do not readily generate ring flips. See Appendix~\ref{app:details} for details.}: both inversion and charge conjugation symmetry need to be broken explicitly. 
%
%

\subsection{Definition of the periodic electric field}
\label{sec:periodicity}
\begin{table}
\begin{tabular}{c|c|c|c}
Object & Position  & Domain\\
\hline
$S^z_j$ & links  & $\{\pm1/2\}$\\
$\nabla_t\cdot S^z$  & tetrahedra & $\{-2,-1,0,1,2\}$\\
$a_j$ & links  & $U(1)$\\
$\nabla\times_p a$ & plaquettes & $U(1)$\\
$e_p$ & plaquettes  & $(-\pi, \pi)$\\
$\nabla\cdot_v  \nabla\times_p a $ & voids  & \{0\}\\
$\nabla\cdot_v  e $ & voids & $2\pi \{-1, 0, 1\}$
\end{tabular}
\caption{Careful definitions of the (classical) coordinate quantities used in QSI (see Fig. \ref{fig:lattice_objects}).}\label{tab:definitions}
\end{table}
%
%
As $a$ is a rotor variable belonging to $U(1)$, $\nabla\times_p a$ is ambiguous up to factors of $2\pi$. Indeed, any summations of $a$ must be understood as an addition of phases, intrinsically modulo $2\pi$. This ambiguity makes it somewhat subtle to define the presence (or absence) of an electric monopole, i.e., a vison. 

There is no ambiguity if one works directly with the Wilson loops $\exp(i \nabla\times_p a)$ (deriving indeed from the product of spin operators around an elementary hexagonal plaquette). Thinking of $a$ as a lattice gauge field, it may be readily observed that this quantity is the total phase acquired by the transport of a virtual $U(1)$ charge around the plaquette $p$, the Wilson loop which contains all the gauge-invariant information of the lattice gauge theory. 

\begin{figure}
\begingroup%
  \makeatletter%
  \providecommand\color[2][]{%
    \errmessage{(Inkscape) Color is used for the text in Inkscape, but the package 'color.sty' is not loaded}%
    \renewcommand\color[2][]{}%
  }%
  \providecommand\transparent[1]{%
    \errmessage{(Inkscape) Transparency is used (non-zero) for the text in Inkscape, but the package 'transparent.sty' is not loaded}%
    \renewcommand\transparent[1]{}%
  }%
  \providecommand\rotatebox[2]{#2}%
  \newcommand*\fsize{\dimexpr\f@size pt\relax}%
  \newcommand*\lineheight[1]{\fontsize{\fsize}{#1\fsize}\selectfont}%
  \ifx\svgwidth\undefined%
    \setlength{\unitlength}{143.00070046bp}%
    \ifx\svgscale\undefined%
      \relax%
    \else%
      \setlength{\unitlength}{\unitlength * \real{\svgscale}}%
    \fi%
  \else%
    \setlength{\unitlength}{\svgwidth}%
  \fi%
  \global\let\svgwidth\undefined%
  \global\let\svgscale\undefined%
  \makeatother%
  \begin{picture}(1,1)%
    \lineheight{1}%
    \setlength\tabcolsep{0pt}%
    \put(0.50961835,0.69076844){\color[rgb]{0.24313725,0.50980392,0.0745098}\makebox(0,0)[lt]{\lineheight{1.25}\smash{\begin{tabular}[t]{l}$p_1$\end{tabular}}}}%
    \put(0,0){\includegraphics[width=\unitlength,page=1]{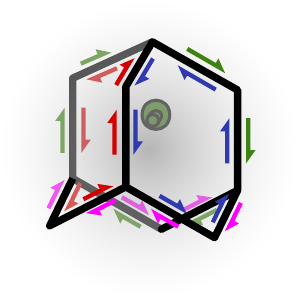}}%
    \put(0.69874631,0.82965421){\makebox(0,0)[lt]{\lineheight{1.25}\smash{\begin{tabular}[t]{l}$a$\end{tabular}}}}%
    \put(0.59840202,0.4985659){\color[rgb]{0.19215686,0.22352941,0.68235294}\makebox(0,0)[lt]{\lineheight{1.25}\smash{\begin{tabular}[t]{l}$p_2$\end{tabular}}}}%
    \put(0.50931649,0.15486541){\color[rgb]{1,0,1}\makebox(0,0)[lt]{\lineheight{1.25}\smash{\begin{tabular}[t]{l}$p_3$\end{tabular}}}}%
    \put(0,0){\includegraphics[width=\unitlength,page=2]{canceling_a.pdf}}%
    \put(0.06536471,0.67514441){\color[rgb]{0.83137255,0,0}\makebox(0,0)[lt]{\lineheight{1.25}\smash{\begin{tabular}[t]{l}$p_4$\end{tabular}}}}%
  \end{picture}%
\endgroup%

\begin{tikzpicture}
\draw (0,0) circle (1);
\filldraw (0,0) circle(0.1);
\begin{scope}[very thick,red, decoration={
    markings,
    mark=at position 0.5 with {\arrow{>}}}
    ] 
    \draw[postaction={decorate}] (0:1)--(30:1) node[midway, right]{$e_0$};
    \draw[postaction={decorate}] (30:1) -- (-45:1) node[midway, below right]{$e_1$};
    \draw[postaction={decorate}] (-45:1) -- (110:1) node[midway, right]{$e_2$};
    \draw[postaction={decorate}] (110:1) -- (0:1) node[midway, above]{$e_3$};
\end{scope}
\draw (0,-1) node[below] {$\nabla\cdot e=0$};
\draw (3,0) circle (1);
\filldraw (3,0) circle(0.1);
\begin{scope}[very thick,red, xshift=3cm, decoration={
    markings,
    mark=at position 0.5 with {\arrow{>}}}
    ] 
    \draw[postaction={decorate}] (0:1)--(30:1) node[midway, right]{$e_0$};
    \draw[postaction={decorate}] (30:1) -- (110:1)node[midway, below]{$e_1$};
    \draw[postaction={decorate}] (110:1) -- (210:1)node[midway, right]{$e_2$};
    \draw[postaction={decorate}] (210:1) -- (0:1)node[midway, below]{$e_3$};
\end{scope}
\draw (3,-1) node[below] {$\nabla\cdot e=2\pi$};
\draw (6,0) circle (1);
\filldraw (6,0) circle(0.1);
\begin{scope}[very thick,red, xshift=6cm, decoration={
    markings,
    mark=at position 0.5 with {\arrow{>}}}
    ] 
    \draw[postaction={decorate}] (0:1)--(160:1)node[midway, right]{$e_0$};
    \draw[postaction={decorate}] (160:1) -- (320:1)node[midway, below left]{$e_1$};
    \draw[postaction={decorate}] (320:1) -- (120:1)node[midway, left]{$e_2$};
    \draw[postaction={decorate}] (120:1) -- (0:1)node[midway, above]{$e_3$};
\end{scope}
\draw (6,-1) node[below] {$\nabla\cdot e= 2\pi$};
\end{tikzpicture}
\caption{
Top panel: Exact vanishing of $\nabla \cdot_v \nabla \times_p a$ on a particular lattice void, composed of the four plaquettes $p_{1,2,3,4}$. One may regard the link variables $a$ as either $U(1)$ rotors or real numbers.
Bottom panel: Summation of phases involved in $V(v)$ (Eq.~\eqref{eq:vison_operator}) as viewed in the $(-\pi, \pi)$ chart.
The topology of $U(1)$ is depicted using a circle, with the four red arrows representing the four $e_p$ fluxes associated with the void $v$. 
As discussed in the main text, the path must return to its starting point to remain consistent with $\nabla\cdot_v \nabla \times_p a = 0 (\rm{mod} 1)$.
The removal of $\pm \pi$ phases forbids any red arrow to enter the midpoint of the circle -- this is in effect a topological singularity about which a winding number can be defined.
}
\label{fig:vison_sketch}
\end{figure}

For each void $v$ of the lattice (see Fig.~\ref{fig:lattice_objects}), we introduce the lattice divergence of $\nabla\times_p a$ to be
\begin{align*}
\nabla \cdot_v (\nabla\times_p a) 
&= \eta_v \sum_{p\in v} (\nabla\times_p a)
\, ,
\end{align*}
where $\eta_v = +1$ on the $A$-sublattice voids and $\eta_v = -1$ on the $B$ voids. It is then tempting to define a vison charge operator $\tilde{V}(v) = ( \nabla\cdot_v \nabla\times a )/(2\pi)$.
Since all $a$ variables are defined modulo $2\pi$, we are forced to consider $\tilde{V}$ modulo $1$. 
Due to the geometric cancellation illustrated in the top panel of Fig.~\ref{fig:vison_sketch}, we have $\tilde{V} \equiv 0\ (\mathrm{mod} 1)$.


%

Nonetheless, it is clearly useful to be able to measure the sense in which the divergence is zero: there is a physical difference, for example, between a void with all plaquette fluxes close to zero and a void with predominantly $\pi/2$ fluxes. In order to construct a vison charge operator that is sensitive to this difference, we have to first choose a chart for the lattice $U(1)$-valued field $\nabla\times a$, i.e., a partial mapping to an open subset of the real axis; we shall refer to $\nabla\times a$ in a given chart as $e_p$, to indicate this subtle but important change. The convention used in the majority of the literature~\cite{szaboSeeingLightVison2019} (implicitly or otherwise) is to take $e_p \in (-\pi,\pi)$, which for weak fields is indistinguishable from standard QED.
Regarding $e_p\in\mathbb{R}$ as $\nabla\times_p a$ in a particular chart, the vison charge operator can then be defined as 
\begin{align}
    V(v) = \frac{1}{2\pi}\eta_v \sum_{p\in v} e_p 
    \, , 
\end{align}
which in general takes any value in $\mathbb{Z}$. This step makes $V$ sensitive to the journey taken by the phase as it loops back to zero, identifying the vison charge as a winding number (bottom panel of Fig.~\ref{fig:vison_sketch}) that, in the chart $(-\pi,\pi)$, may only take values $-1,0,1$. 

For consistency with the existing QSI literature, we will also work in the chart-fixed `0-flux convention' in which $e\in (-\pi,\pi)$, leaving further comments on chart dependence to Appendix~\ref{app:chart_indep}. As a manifestly chart-dependent quantity, $V$ should be interpreted as a vison marker, rather than the exact vison charge operator itself. Where appropriate, we will also take steps (see Appendix~\ref{app:chart_indep}) to illustrate that any electric sector behavior we remark on is also observed in terms of purely chart-independent variables.
%
%

\subsection{Quadratic Hamiltonian
\label{ssec:quadratic}}
In the special cases $(g,g') = (\pm 1, 0)$ and $(g,g') = (0,\pm 1)$, the ground state of the semiclassical Hamiltonian in Eq.~\eqref{eq:large_S_ham} can be read off as $\nabla\times_p a = 0, \pi, \pm\pi/2$, $\forall\,p$, which are the four uniform states consistent with the constraint $\nabla\cdot_v \nabla\times_p a = 0 \: (\text{mod} \, 2\pi)$. We name these states $U(1)_0$, $U(1)_\pi$ and $U(1)_{\pm \pi/2}$, respectively. Using these four states as variational Ans\"{a}tze, one finds that 
\begin{equation}
    \rho_g \operatorname{min}_{n\in\mathbb{Z}_4} 
    \cos\left(\frac{\pi n}{2}-\phi_g\right)
    \label{eq:upperbound}
\end{equation}
is an upper bound on the ground-state energy.

These uniform-flux states are robust against perturbations. Indeed, if we write the Hamiltonian~\eqref{eq:large_S_ham} in terms of $e_p$ (upon fixing a chart) and we consider small fluctuations $e_p \ll \pi/2$, 
\begin{align}
    H &\sim \sum_{p} \left[ -g -g' e_{p}+ \frac{g}{2}e_{p}^2  + O(e_p^3) \right]\\
     &= -4gN -g' \sum_{v\in A}\nabla \cdot_v e_p + \sum_{p} \frac{g}{2}e_{p}^2 + O(e_p^3)
     \, , 
     \label{eq:quadratic_guess}
\end{align}
it becomes clear that in the vison-free $\nabla \cdot_v e_p = 0$ sector (as appropriate for the low energy $U(1)_{0}$ state), the $g'$ term contributes only at higher than quadratic order. 

Note that the divergence term, 
\begin{align}
    \nabla\cdot_v e_p \equiv \eta_v \sum_{p\in v} e_{p} = 2\pi V(v)\ ,
    \hspace{1em}\eta_v = \pm 1\text{ for }v\in A/B 
    \, ,
    \label{eq:vison_operator}
\end{align}
is a multiple of the large-$S$ vison charge operator $V(v)$.
The sum appearing in Eq.~\eqref{eq:quadratic_guess} runs over $A$ voids only, and we can thus identify the linear-order term as a staggered electric potential that favors positive visons on the $A$ voids (and therefore negative ones on the $B$ voids, by global charge neutrality). Due to the vison quantization, there is no linear response to a small $g'$ perturbation of the $U(1)_{0}$ phase, up until $g'$ reaches a value comparable to the vison gap. Similarly, one can show that the $U(1)_{\pm\pi/2}$ phases have vanishing linear response to $g$ perturbations. 

To measure the excess of positive visons on the $A$ sublattice, we define the order parameter as the average value of the vison charge operator restricted to voids lying on the $A$ sublattice: 
\begin{align}
V_A 
= \sum_{v\in A} \frac{V(v)}{N_{\mathrm{tetra}}} 
= \sum_{p} 
  \frac{e_p}{2\pi N_{\mathrm{tetra}}}
\, , 
\end{align}
which has vanishing expectation value at all temperatures when $g'=0$, and tends to $1$ at low temperatures in the vison crystal phase. This quantity essentially measures the electric AIAO polarization (recall that by global vison charge neutrality, $V_B = -V_A$), playing the role of the staggered magnetization for a N\'{e}el antiferromagnet. 
%
%

\subsection{Enhanced spectral periodicity and duality in the minimal model
\label{sec:duality}}
\begin{figure}
\centering
\includegraphics[width=0.2\textwidth]{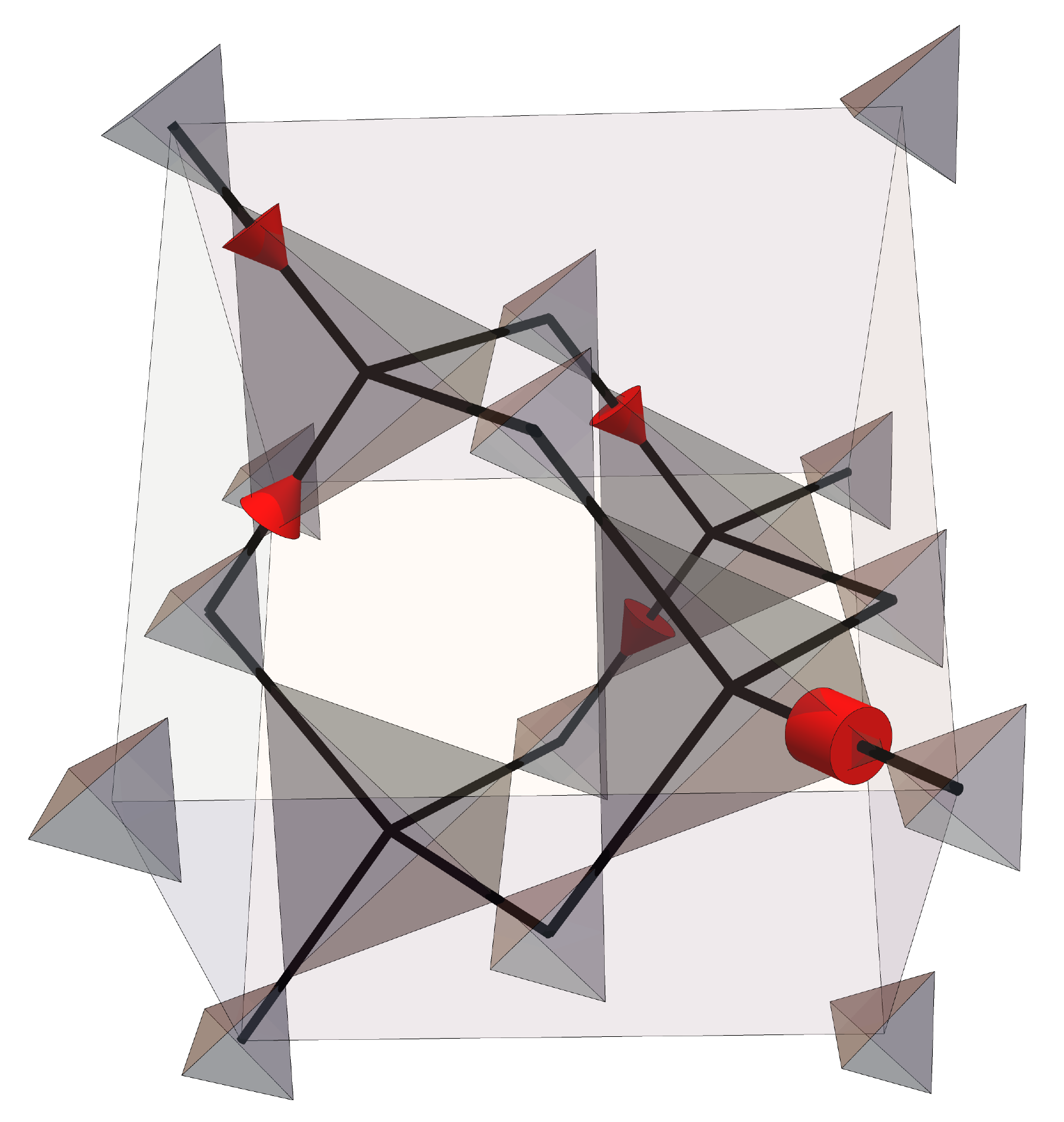}
\includegraphics[width=0.2\textwidth]{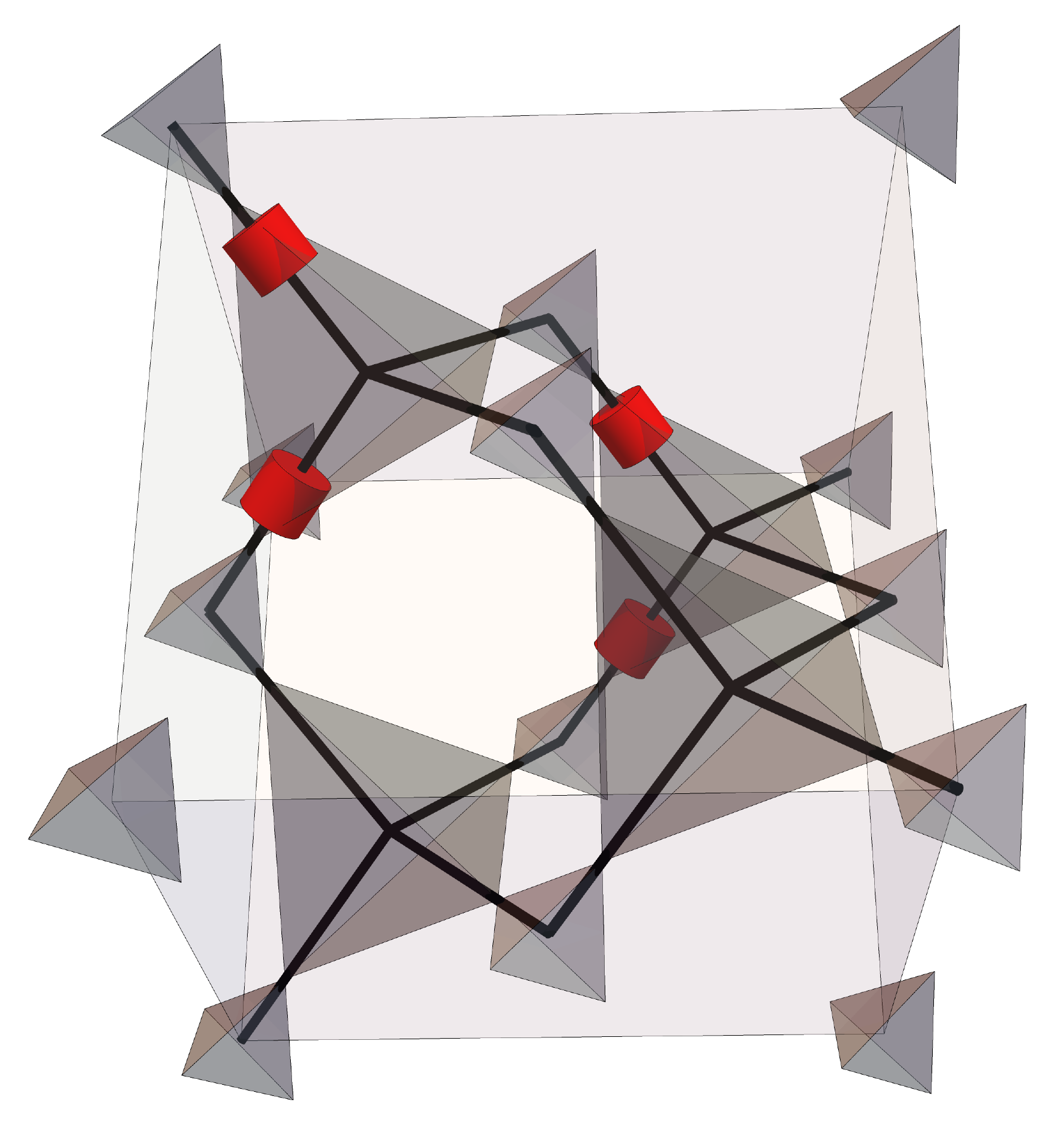}
\caption{Explicit gauge field configurations realising uniform flux phases. Left panel: $a_{\pi/2}$, generating the $U(1)_{\pi/2}$ phase. Right panel: $a_\pi$, generating the $U(1)_\pi$ phase. Red arrows indicate gauge field values of $a = \pi/2$ (positive in the direction of the arrow); red cylinders indicate $a=\pi$ (no direction needed as $+\pi = -\pi$); all unlabeled bonds have $a=0$. These examples were chosen such that on every link $j$, $2a_{\pi/2}(j) = a_\pi(j)$.} 
\label{fig:duality_map}
\end{figure}

Examples of gauge field configurations for the $U(1)$ lattice gauge theory on the diamond lattice, generating uniform $\pi/2$ and $\pi$ flux phases are given in Fig.~\ref{fig:duality_map}. 
Though the fluxes themselves are commensurate with the periodicity of the breathing lattice, the enlargement of the unit cell retains a physical significance in the periodicity of the spinon continuum~\cite{chenSpectralPeriodicitySpinon2017, desrochersSymmetryFractionalizationGauge2023}. In essence, the Aharonov-Bohm phase acquired from spinon transport fractionalizes the lattice translational symmetry into a projective space group representation, which only `squares out' to a true representation when an enlarged unit cell is considered. The zero-flux phase can be described by tiling a primitive unit cell with $a_{rr'} = 0$ over the whole lattice, while the smallest link variable configuration of a $\pi$ flux phase requires two primitive unit cells. It can be verified explicitly that the smallest possible unit cell describing a $U(1)_{\pi/2}$ phase consists of four primitive unit cells, distinguishing the state from the $U(1)_0$ and $U(1)_\pi$ phases.

Fig.~\ref{fig:duality_map} may also be interpreted as a duality of the effective Hamiltonian. This is achieved by recognising that the Abelian group structure of the lattice gauge theory extends to the spin model. If $a(j)$ is a lattice gauge field, then the unitary operation on $\#$ 
\begin{align}
    \mathcal{R}_a[ \# ] = \left(\prod_{j \in \{\text{links}\}} e^{i a(j)S^z_j} \right)\# \left( \prod_{j' \in \{\text{links}\}} e^{-i a(j')S^z_{j'}}\right)
    \label{eq:R_definition}
\end{align}
represents a transformation acting on the spin model which rotates the phase of the ring flip operator on any plaquette $p$ by $\mathcal{O}_p \mapsto e^{i\nabla\times_p a} \mathcal{O}_p$. The $\pi$-flux configuration $a_\pi(j)$
of Fig.~\ref{fig:duality_map} (right panel) results in $\mathcal{R}_{a_\pi} [\mathcal{O}] =  -\mathcal{O}$, reproducing the well known  duality~\cite{leeGenericQuantumSpin2012,chenSpectralPeriodicitySpinon2017} between the $U(1)_0$ and $U(1)_\pi$ phases. The left panel of the same figure is a
`square root' of the $U(1)_\pi \leftrightarrow U(1)_0$ transformation, generating $\mathcal{R}_{a_{\pi/2}}[\mathcal{O}_p] = i \mathcal{O}_p$ for all $16$ plaquettes in the unit cell. 

Charge conjugation, combined with the $\pi/2$ rotation duality, makes the physics of the $g,g'$
phase diagram $D_4$ symmetric. The $g,g'$ model can therefore be fully characterized by the half-quadrant $g>0, 0<g'<g$. In the special cases $\phi_g \in \frac{\pi}{4} \mathbb{Z}_8$, these dualities are elevated to $\mathbb{Z}_2$ symmetries of the Hamiltonian, which we further distinguish into the `axis' symmetry lines $g'=0$, $g=0$ and the `diagonal' symmetry lines $g=\pm g'$. While the axis symmetries remain intact down to zero temperature, we will show that the diagonal symmetries are spontaneously broken below a critical temperature.
%
%

\section{Large-$S$ phase diagram
\label{sec:phasedia}}
We are finally in the position to study the phase diagram of the model in the large-$S$ limit. We use a semiclassical Monte Carlo approach developed by Szab\'{o} and Castelnovo~\cite{szaboSeeingLightVison2019}, based on the large-$S$ expansion~\cite{kwasigrochSemiclassicalApproachQuantum2017}, adapted to our effective quantum Hamiltonian in Eq.~\eqref{eq:heff}. 
%
We perform classical Monte Carlo on the low-energy effective Hamiltonian (Eq.~\ref{eq:heff}), capturing the gauge-sector dynamics of the quantum theory in a large-$S$ sense~\footnote{The reader is referred to Appendix~\ref{app:numerics} and to Ref.~\onlinecite{szaboSeeingLightVison2019} for further details.}.
The ring flip operator $\mathcal{O}_p = S^+S^-S^+S^-S^+S^-$ is regarded as a complex number of norm $\le 1$. The presence of a vison in void $v$ is measured by $V(v) = \sum_{p \in v} \arg \mathcal{O}_p /2\pi$, using the chart $\arg : \mathbb{C} \to [-\pi, \pi)$.

We initialize the system in a random state, then
lower the temperature from $T=10\rho_g$ to $T=10^{-5}\rho_g$ in logarithmically spaced steps. At every temperature, we run a number of annealing steps known from benchmarks to equilibrate the system (typically ranging from 128 at high temperature to 512 at low temperature). At the end of a full cooling cycle, we then re-heat the system up to the initial temperature, following the same temperature steps and comparing order parameters throughout. The absence of any hysteresis is used as indication that equilibrium has been likely attained in the simulations.

\begin{figure}
\includegraphics{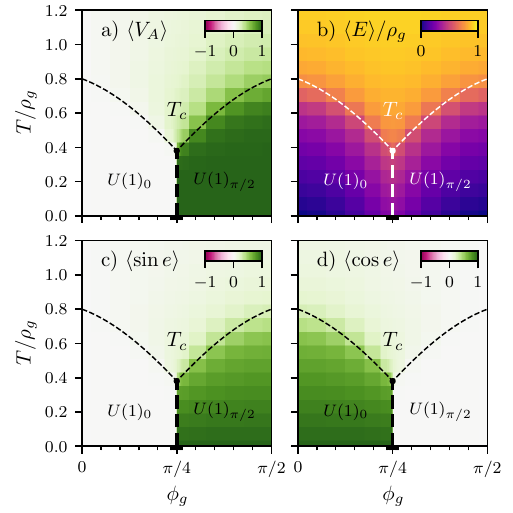}
\caption{\label{fig:phasedia} 
Phase diagram of the $g+ig' = \rho_g \exp(i\phi_g)$ model from semiclassical numerics, showing a) the average expectation value of the large-$S$ average vison charge $V_A$ on $A$ tetrahedra, b) the energy of the system relative to the 0-flux state, c) and d) the expectation values $\langle \sin e \rangle = \sum_p \sin e_p / N_p$ and $\langle \cos e \rangle = \sum_p \cos e_p / N_p$, where $N_p$ denotes the total number of plaquettes.
The black, dotted curve $T_{pm}/ T_c = \sqrt{2} \left[1.3\operatorname{max}(|\cos(\phi_g)|, |\sin(\phi_g)|) -0.3\operatorname{min}(|\cos(\phi_g)|, |\sin(\phi_g)|)\right]$, is given as a guide to the eye, outlining the crossover between paramagnetic and $U(1)$ behaviour. The resolution of $\phi_g$ is $\sim \pi/100$ in the region $|\phi_g-\frac{\pi}{4}| < 0.1$, and $\pi/19$ elsewhere, with data points indicated by the x-axis ticks.
}
\end{figure}
A sharp phase transition is observed in Fig.~\ref{fig:phasedia} at $g=g'$ at low temperatures, separating the conventional $U(1)_0$ phase from a phase in which all $A$ sites are occupied by visons. In the latter, all electric fluxes take the value $\pi/2$, hence the name $U(1)_{\pi/2}$ phase, previously identified using a projective symmetry group approach in Ref.~\onlinecite{desrochersSymmetryFractionalizationGauge2023}.

The transition from 0-flux to $\pi/2$-flux phases may be understood as the chemical forcing $g'$ overwhelming the intrinsic vison chemical potential $\mu_{g'=0} = 7.872367608(68) g$~\cite{szaboSeeingLightVison2019}. Neglecting any cohesive energy from interactions, the phase transition would occur at $g'/g = \mu_{g'=0}/g \simeq 7.872$ ($\phi_g \simeq 0.460\pi$). However, we know from the (unitary) duality operation $\mathcal{R}_{a_{\pi/2}}\mathcal{C}$, defined in 
Eqs.~\eqref{eq:R_definition} and~\eqref{eq:C_definition}, 
that the phase-boundary structure of the global phase diagram must be symmetric under $g \leftrightarrow g'$.
It follows that the critical $g'/g$ is at most $1$ (since $g/g'$ is also a critical point).
Our numerics clearly show $g'/g=1$ to be the critical point, emphasizing the substantial renormalization of short-ranged $\mu$ by interactions.

We verified that the zero-temperature system energy saturates the trivial bound in Eq.~\eqref{eq:upperbound} from Sec.~\ref{ssec:quadratic} to within Monte Carlo error, suggesting that we have successfully identified all relevant low temperature phases. Further, it suggests that the smaller of $g, g'$ plays a trivial role at zero temperature, merely shifting the ordinary QSI vacuum.

Fixing the overall energy scale $\rho_g$, we identify a characteristic temperature $T_{pm}$ associated with the crossover to the disordered phase. Guided by the quadratic theory of Sec.~\ref{ssec:quadratic}, we are able to establish a phenomenological model that captures this behaviour.
Recall that expanding the Hamiltonian about the 0-flux vacuum, Eq.~\eqref{eq:quadratic_guess}, yields $g$ as the tension of the electric field (i.e., the permittivity of free space) and $g'$ as a staggered chemical potential for visons. Applying the duality $\mathcal{R}_{a_{\pi/2}}$ interchanges $g$ and $g'$, rendering $g'$ the effective field tension for excitations above the $\pi/2$-flux state, which is indeed the preferred ground-state flux configuration when $g' > |g|$.
It may be seen by a straightforward generalization that the quantities $\mathcal{T} = \max(|g|,|g'|)$ and $U_s = \min(|g|,|g'|)$ act respectively as field tension and chemical potential for all values of $\phi_g$.
We see that $T_{pm} \sim \sqrt{2}T_c (1.3 \mathcal{T}  -0.3 U_s )$ captures well the general trend of the crossover. Though the dominant contribution to the crossover's shape is the weakening of field tension, including a small $U_{s}$ contribution appears to improve the visual agreement. 

Raising $T$ above $T_{pm}$ brings about a vison liquid state, featuring a random disordered electric field. We identify this with spinon-free classical spin ice. The transition to the true paramagnetic phase~\cite{gingrasQuantumSpinIce2014} is not observable in our approximation scheme, as we work within the no-spinon ensemble of states.



%
%

\subsection{Return to microscopic coupling constants}
\begin{figure}
    \includegraphics{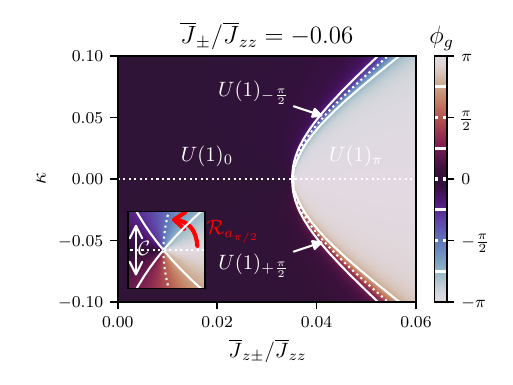}
    
    \caption{AIAO bias $\phi_g$ as a function of the uniform anisotropy parameter $\kappa = J_{zz,A}/J_{zz,B}=J_{\pm,A}/J_{\pm,B}=J_{z\pm,A}/J_{z\pm,B}$ and of $\overline{J}_{z\pm}/\overline{J}_{zz}$. Sublattice-averaged $\overline{J}_{\pm}, \overline{J}_{zz}$ are defined in the obvious way. Dotted contours represent lines of $g'=0$ and $g=0$. Solid lines represent zero-temperature phase boundaries, as established by the numerics (see Fig.~\ref{fig:phasedia}). Inset: detail near the singular point $g=g'=0$, annotated with the two generating dualities of the model, charge conjugation $\mathcal{C} : g'\mapsto -g'$ and onsite rotation $\mathcal{R}_{\pi/2} : g\mapsto g', g'\mapsto -g$.}
    \label{fig:parameter_control}
\end{figure}

It is useful to now reconsider the phase diagram in terms of the original microscopic variables of Sec.~\ref{sec:model}, $J_{zz}$, $J_\pm$, and $J_{z\pm}$. A particular slice of parameter space is shown in Fig.~\ref{fig:parameter_control}, using the uniform scaling parametrization of Sec.~\ref{sec:perturb}. Close to the critical point $\kappa=1$, $\bar{J}_{z\pm}/\bar{J}_{zz} -\sim 0.2$, it is possible to use $\overline{J}_{z\pm}/{\overline{J}_{zz}}$ and $\frac{\kappa-1}{\kappa+1}$ as proxies for $g,g'$, see Eqs.~\eqref{eq:g_linearised} and~\eqref{eq:gp_linearised}. We thereby obtain a clear physical interpretation of the $D_4$ duality structure derived in Sec.~\ref{sec:duality} -- the four white lines (two dotted, two solid) are lines of emergent symmetry generated by the conjugate-shift operations $\mathcal{R}_{a_{\pi/2}}^n\mathcal{C}$, $n\in\mathbb{Z}_4$.

Admittedly, the value of $\overline{J}_{z\pm}$ needed for $g'$ to overcome $g$ is very large compared with $\overline{J}_\perp$. There is therefore a risk of the Ising interactions $I_2$ causing the system to order before any appreciable $g'$ can be generated. We stress, however, that the calculation here is truncated at fourth order. We expect very substantial renormalization of $g$ and $g'$ at higher orders. 
%
%

\subsection{Critical end point}
The duality arguments of Sec.~\ref{sec:duality} are consistent with a continuous connection between the $U(1)_0$ and $U(1)_{\pi/2}$ phases via the phase-incoherent spin ice paramagnetic phase -- there is no symmetry breaking across the transition. The line of first order phase transitions at $g'/g=1$ therefore ends at a critical end point. 

We probed the critical exponents of the end point using finite size scaling analysis.
We used 3D periodic systems of cubic side lengths $L=4-12$ unit cells. At each system size, we initialized $256$ independent systems to random initial states, then performed a temperature sweep from $10\rho_g$ down to $0.3\rho_g$, then back up to $0.45\rho_g$. Annealing times were chosen to remove any hysteresis in the observables. In the case of specific heat, we computed an ad hoc cost function~\footnote{In this case, the binned variance described in Ref.~\onlinecite{newmanMonteCarloMethods1999}.} designed to estimate the quality of the scaling collapse over a grid of $T_c$, $\alpha$ and $\nu$ values. From this tableaux, we estimated both the optimal fitting parameters and their uncertainties, including $T_c$, based on the width of the optimal fit basin. We repeated this analysis for the staggered polarization data $\langle V_A \rangle$, with the additional fitting parameter $\langle V_A\rangle_c$, obtaining also values for $\beta$ and $\gamma$. 

\begin{figure}
\includegraphics{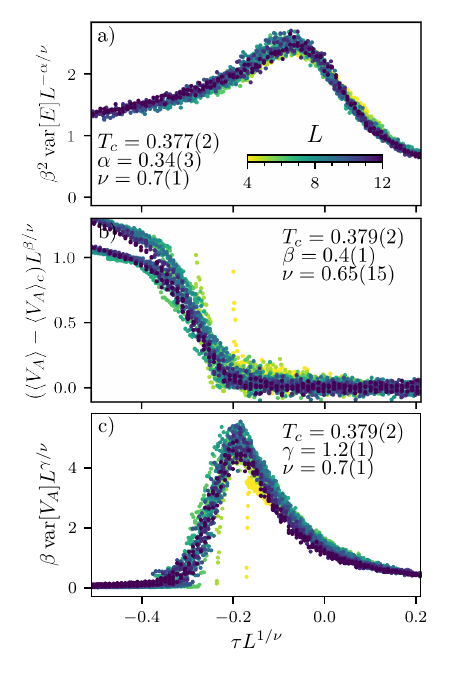}
\caption{Finite-size scaling collapse at the critical end point of the $g=g'$ line, in terms of $\tau = (T-T_c)/T_c$, for systems consisting of $L^3$ cubic unit cells. 
a) Specific heat capacity, $\text{var}[E]$. 
b) Reduced thermally averaged staggered vison order parameter, $\langle V_A \rangle - \langle V_A \rangle_c$, where $\langle...\rangle_c$ denotes the thermal average at the critical point.
c) Susceptibility of the order parameter, $\text{var}[V_A]$. 
We performed a fit for $T_c$ and the relevant scaling exponent for each of the three quantities separately, with results given directly in each panel. 
For reference, the critical exponents of the 3D Ising model are $\nu = 0.629971(4), \alpha = 0.11008(1), \beta = 0.326419(3), \gamma=1.237075(10) $~\cite{kosPrecisionIslandsIsing2016,reehorstRigorousBoundsIrrelevant2022}.}
\label{fig:criticality}
\end{figure}

Our results (shown in Fig.~\ref{fig:criticality}) are suggestive of 3D Ising criticality, finding its critical exponents $\beta$, $\gamma$ and $\nu$ to lie within our error bars.
%
Recall from Sec.~\ref{sec:duality} that on the line $\phi_g=\pi/4$, the operator $\mathcal{R}_{a_{\pi/2}} \mathcal{C}$ is precisely a $\mathbb{Z}_2$ symmetry that is spontaneously broken across the $g'=g$ line -- consistently with Ising criticality at the end point. The story is, however, not quite so neat: the exponent $\alpha$ overestimates the Ising value by a factor of $3$, well beyond what can be explained by our fitting uncertainty. It shall remain an intriguing open question for future work. 

This peculiarity is perhaps less surprising against the background of Debye-H\"{u}ckel plasma criticality, which has been variously reported as Ising, Gaussian, crossover-like and tricritical~\cite{fisherStoryCoulombicCritiality1994, levinCoulombicCriticalityGeneral1994, fisherGinzburgCriterionCoulombic1996}, depending sensitively on the form of the short-range cutoff~\cite{brilliantovPeculiarityCoulombicCriticality1998}. 
%
%

\subsection{Compactness of the $e$ field.}
It is worth remarking on the aspects of the phase diagram that are not symmetric about $\phi_g = \pi/4$. 
Physically, the vison charge on $A$ tetrahedra $V_A = \sum_{v \in A} \nabla \cdot_v e_p / ( 2 \pi N_{\text{void}} )$ is a multiple of the expectation value of the electric field $\frac{1}{N} \sum_p e_p/(2\pi)$, which one would naively expect to average exactly to $1/2$ on the phase boundary between the $U(1)_0$ and $U(1)_\pi$ phases above $T_c$.
However, the critical value of the vison order parameter $\langle V_A \rangle_c \simeq 0.38$, and perhaps even more surprisingly the high-temperature phase-incoherent state sees an even smaller value, $\langle V_A \rangle \simeq 0$.

To understand this, let us firstly note that systems with $\phi_g = x$ and $\phi_g = \pi/2 - x$ are dual to one another. Indeed, the system energy seen in Fig.~\ref{fig:phasedia}b is symmetric about the coexistence line $\phi_g = \pi/4$ to within simulation accuracy, and the expectation values $\langle \cos e \rangle$ and $\langle \sin e \rangle$ in Fig.~\ref{fig:phasedia}c,d are related to one another by mirroring about the coexistence line.

Numerically, the order parameter $V_A$ is computed by $V = \sum_{p} \mathrm{arg}(\mathcal{O}_p)$, where $\mathcal{O}_p$ is a complex number formed by a product of $S^x \pm i S^y$ terms using the chart $\operatorname{arg}(z) \in (-\pi, \pi]$. As already discussed in Sec.~\ref{sec:periodicity}, this is a choice of chart convention that privileges zero, as at high temperature, the system samples a uniform distribution on $(-\pi,\pi]$. This average contains no physical information, only a marker of which chart was used for $e_p$.
As the temperature is raised, $\langle V_A \rangle$ `crosses over' from a useful indicator of the number of topological defects to a completely unphysical constant. This emphasizes the ill-definedness of visons in the high density regime~\cite{luscherTopologyAxialAnomaly1999}. For further discussion of the precise meaning of the vison in QSI lattice gauge theory, see Appendix~\ref{app:chart_indep}. 
%
%

\section{Electrodynamics at finite $g'$}
\label{sec:uniform}
\begin{figure}
\includegraphics{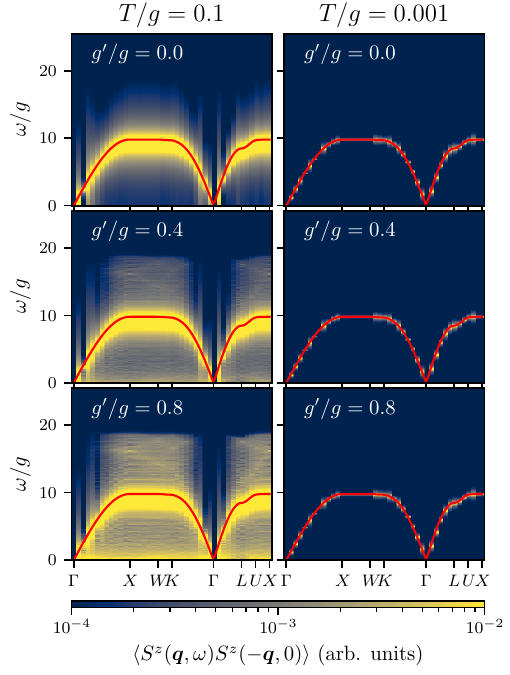}
\caption{Dynamical structure factor $\langle S^z(\boldsymbol{q}, \omega) S^z(-\boldsymbol{q}, 0) \rangle$ at constant $g$, with $T/g=0.1$ (left panels) and $T/g=0.001$ (right panels), for various values of the ratio $g'/g$ (in the $U(1)_{0}$ phase, equivalent to $U(1)_{\pi/2}$ by symmetry). The solid red lines show the analytic (large-$S$) photon dispersion from Ref.~\onlinecite{bentonSeeingLightExperimental2012}. The system size is $L=12$.
}
\label{fig:dsf_all}
\end{figure}

The dynamical structure factor of our system (see Fig.~\ref{fig:dsf_all}) continues to feature the known large-$S$ photon dispersion in the presence of $g'$.  
We interpret these plots as probes of the photon spectrum alone. The visons possess no kinetic properties in the large-$S$ limit and only hop via instanton processes; we are therefore unable to meaningfully probe their real-time evolution~\cite{szaboSeeingLightVison2019}. For this reason, our simulations show a gapless photon dispersion even when temperatures are high enough to have a substantial vison population~\cite{szaboSeeingLightVison2019}, as opposed to the Debye-screened mass one would expect from a charged plasma~\cite{kwasigrochVisongeneratedPhotonMass2020, 
bentonSeeingLightExperimental2012}. 

The $\langle S^z S^z \rangle$ correlations are invariant under the $\mathcal{R}_{a_{\pi/2}}$ duality transformation, and therefore the behavior in the $U(1)_{\pi/2}$ phase is identical to that shown in Fig.~\ref{fig:dsf_all} (with the roles of $g$ and $g'$ reversed). This statement was verified numerically as a benchmark.

At $g'=0$ we see that the familiar large-$S$ photon dispersion is in excellent agreement with the analytic result at low temperature. Less obviously, the vast majority of the spectral weight remains identical to the $g'=0$ large-$S$ photon curve even when $g'/g$ is of order $1$. It is only at higher temperatures, where the photon states become highly populated, that the $e_p^3$ vertex -- see Eq.~\eqref{eq:e3_vertex} -- begins to have a measurable influence. In this regime we observe a continuum above the photon line that terminates abruptly at twice the renormalized maximum dispersion energy. There is some structure discernible in the continuum, which reflects the density of two-photon states. Excitingly, these anomalous features are visible in the experimentally accessible $S^z$ correlation function, and could in principle be used to detect a system sufficiently close to the phase transition between the $U(1)_{0}$ and $U(1)_{\pi/2}$ phases, at intermediate temperatures. 

The leading order contribution due to the $g'$ term in the $U(1)_0$ phase is cubic, which we suggest is analogous to strong light-matter coupling in a nonlinear crystal. From this perspective, the high-energy features are naturally interpreted as a two-photon continuum arising from three photon up-conversion~\cite{porterFluorescenceExcitationAbsorption1961}, which we compute analytically in the large-$S$ limit for comparison hereafter.
%
%

\subsection{Large-$S$ calculation of two-photon continuum}
\label{ssec:largeS-main}
We show here how the continuum seen in the $\langle S^z S^z \rangle$ dynamical structure factor in Fig.~\ref{fig:dsf_all} can be explained within perturbative large-$S$ field theory. We perform the calculation for the $U(1)_0$ phase, noting that the calculation would be identical in the $U(1)_{\pi/2}$ or $U(1)_\pi$ phases, up to duality relations. 

We decompose the large-$S$ Hamiltonian into quadratic and interacting parts in the standard fashion: 
\begin{align}
H_0 &= \sum_p \frac{g}{2}e_p^2 + \sum_v g' 2\pi V(v)\\
H_I &= \sum_p -\frac{g'}{3!}e_p^3 - \frac{g}{4!}e_p^4 +\frac{g'}{5!} e_p^5+ \frac{g}{6!}e_p^6 + ... \label{eq:e3_vertex}
\, . 
\end{align}
The key insight is that the vison contribution $V(v)$ is a non-dynamical, topological term that does not enter the perturbation theory.

To pick out the most relevant terms from this expansion, we follow Ref.~\onlinecite{kwasigrochSemiclassicalApproachQuantum2017} and perform naive RG scaling on $(\nabla\times_p a)^n$:
\begin{align*}
    \mathbf{r},\, \tau &\mapsto b\mathbf{r},\, b\tilde{s}\tau\\
    a &\mapsto b^{\frac{1-d}{2}} \tilde{s}^{-\frac{1}{2}} a\\
    \int d\tau\,d^{d}\mathbf{r}\, (\nabla\times_p a)^n &\mapsto
    \left(b^{d+1} \tilde{s}\right)^{1-\frac{n}{2}}
    \int d\tau\, d^{d}\mathbf{r}\, (\nabla\times_p a)^n
    \, ,
\end{align*}
where $b>1$. In particular, we note that a pair of contracted three-photon vertices scales as $b^{-1-d} \tilde{s}^{-1}$, which is the same as a Hartree-Fock vertex.
It follows that we need only consider the (momentum-space) diagrams a), b) and c) shown in Fig.~\ref{fig:feyn_diag}:
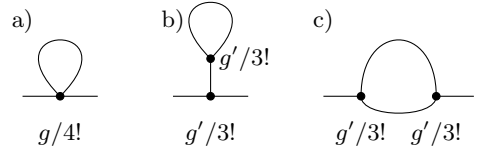
\begin{figure}[h!]
\begin{tikzpicture}
    \draw  (0,0) coordinate (O4)
    .. controls ++(1,1) and ++(-1,1) ..
    (O4) ++(0,-0.5) node{$g/4!$} (O4) -- ++(0.5,0) (O4)--++(-0.5,0);
    \draw (3.5,0) -- ++(0.5,0) coordinate(A) .. controls ++(0,1) and ++(0,1) .. ++(1,0) coordinate(B) --++(0.5,0)
    (A)++(0,-0.5) node{$g'/3!$} 
    (A).. controls ++(0,-0.3) and ++(0,-0.3) .. ++(1,0) (B)
    ++(0,-0.5) node{$g'/3!$} ;
    \draw (1.5,0) --++(0.5,0) coordinate (X) ++(0,-0.5) node{$g'/3!$} (X) --++(0.5,0)
    (X) -- ++(0,0.5) .. controls ++(1,1) and ++(-1,1) .. ++(0,0) coordinate (Y) node[right]{$g'/3!$};
    \filldraw 
    (A) circle(0.05)
    (B) circle(0.05)
    (X) circle(0.05)
    (Y) circle(0.05)
    (O4) circle(0.05);
    \draw
    (O4)++(-0.5,1) node {a)}
    (A)++(-0.5,1) node {c)}
    (X)++(-0.5,1) node {b)};
\end{tikzpicture}
\caption{Leading-order loop corrections in the large-$S$ field theory.}
\label{fig:feyn_diag}
\end{figure}

The Hartree-Fock contribution a) is known~\cite{kwasigrochSemiclassicalApproachQuantum2017} to be a renormalization of the photon continuum. The contribution b) can be shown to vanish exactly because the central propagator carries zero momentum. Remarkably, only the leading-order non-trivial term c) is found to give a contribution from $g'$ to the spectral weight (see Appendix~\ref{app:continuum_calculation} for details). 
\begin{figure}
    \includegraphics{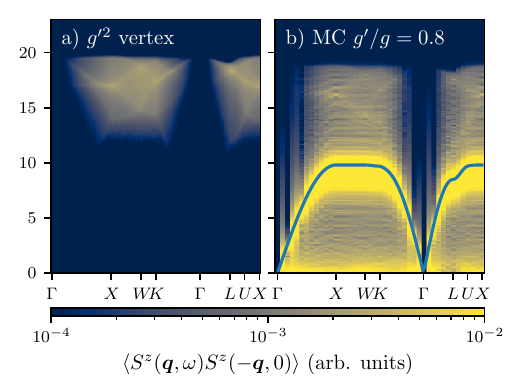}
    \caption{Two-photon continuum in the $g,g'$ model. a) Contribution from the $g'^2$ term to the self energy (arb. intensity units), calculated analytically as discussed in the main text and Appendix~\ref{app:continuum_calculation}. b) Monte Carlo (MC) numerical results from Fig.~\ref{fig:dsf_all}, for $g'/g=0.8$. Both panels correspond to $T = 0.1g$.}
    \label{fig:upconversion}
\end{figure}
Looking at Fig.~\ref{fig:upconversion}, we see excellent agreement between analytical and numerical results. The leading-order calculation somewhat overestimates the maximum energy of the continuum, and in particular the spectral weight at high energy, we expect however that such discrepancies will be resolved by dressing the propagator at higher orders in perturbation theory. 
%
%

\subsection{Static Structure Factor}
\label{sec:ssf}
\begin{figure*}
    \includegraphics{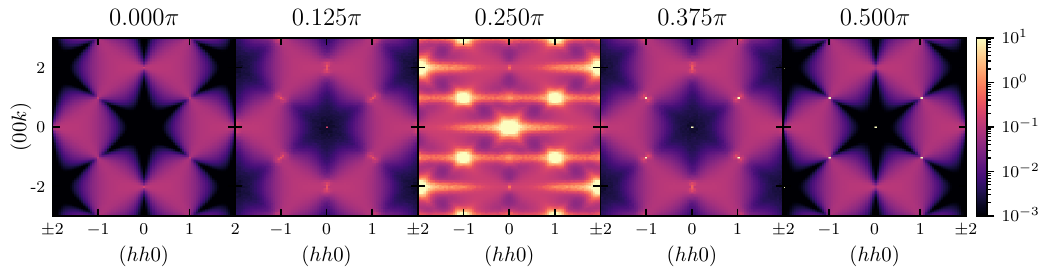}
    \caption{Evolution of the static electric field correlator $\langle e(\boldsymbol{k},0)e(-\boldsymbol{k},0) \rangle$ for a 128,000 site system, across several values of $\phi_g=0, \pi/8, \pi/4, 3\pi/8$ and $\pi/2$ (from left to right), while the ratio $T/\rho_g=1/2$ is held constant. This trajectory in $g,g'$ space moves from deep in the $U(1)_0$ phase to deep in the $U(1)_{\pi/2}$ phase via the disordered state, crossing the coexistence point $\phi_g=\pi/4$ above $T_c$.
    Narrow peaks within the pinch points $(1,1,\pm1)$ and $(-1,-1,\pm1)$ are numerically equal, but may appear inequivalent due to plot rasterization. See Fig.~\ref{fig:ssf_cos_sin} in Appendix~\ref{app:ssf_chart_indep} for a chart-invariant version of this plot.}
    \label{fig:ssf}
\end{figure*}

Perhaps the most direct measurement of vison crystal formation is the appearance of Bragg peaks in the $\langle e e \rangle$ correlations, shown in Fig.~\ref{fig:ssf}. 
Starting from $\phi_g=0$, screening of the Coulomb interaction by the Debye plasma brings about a Lorentzian blurring of the $\langle e e \rangle$ pinch points. This crossover is controlled by two factors: the lowering of the vison gap upon increasing $g'$, and the simultaneous increase of the ratio $T/g$ (recall that we are keeping the ratio $T/\rho_g$ fixed at $1/2$). Both effects independently lead to an increase in the vison density. Interestingly, we observe a blurring of the pinch points even if $g'$ is increased at constant $T/g$ (likely due to charge disorder upon approaching coexistence at the transition line).

On either side of the coexistence line $\phi_g=\pi/4$, a Bragg peak emerges at the $\Gamma$ point due to the statistically significant excess of net vison charge on the $A$ sites (as already mentioned in preceding sections, $e(k=0)$ is indeed proportional to the vison order parameter $V_A$). 

At the coexistence point $\phi_g = \pi/4$, we observe rod correlations associated with domain walls that are frozen in as they become metastable at low temperature. Rods extend in the [110] and (to a lesser extent) in the [111] directions, reflecting an anisotropic energy cost to domain wall orientation.
%
%
%

\section{Vison interactions}
\label{sec:vison_interactions}
Finally, we study the energetics of the visons in our system -- namely their chemical potential and interactions -- within the large-$S$ approximation. 
In order to do so, we create spin configurations where we arrange visons in a polarization-free octupole pattern and measure the total field energy at vanishing temperature, which we then decompose into a chemical potential part and an interaction part by changing system size. 

\begin{figure}
\begingroup%
  \makeatletter%
  \providecommand\color[2][]{%
    \errmessage{(Inkscape) Color is used for the text in Inkscape, but the package 'color.sty' is not loaded}%
    \renewcommand\color[2][]{}%
  }%
  \providecommand\transparent[1]{%
    \errmessage{(Inkscape) Transparency is used (non-zero) for the text in Inkscape, but the package 'transparent.sty' is not loaded}%
    \renewcommand\transparent[1]{}%
  }%
  \providecommand\rotatebox[2]{#2}%
  \newcommand*\fsize{\dimexpr\f@size pt\relax}%
  \newcommand*\lineheight[1]{\fontsize{\fsize}{#1\fsize}\selectfont}%
  \ifx\svgwidth\undefined%
    \setlength{\unitlength}{219.13677185bp}%
    \ifx\svgscale\undefined%
      \relax%
    \else%
      \setlength{\unitlength}{\unitlength * \real{\svgscale}}%
    \fi%
  \else%
    \setlength{\unitlength}{\svgwidth}%
  \fi%
  \global\let\svgwidth\undefined%
  \global\let\svgscale\undefined%
  \makeatother%
  \begin{picture}(1,0.58577543)%
    \lineheight{1}%
    \setlength\tabcolsep{0pt}%
    \put(0,0){\includegraphics[width=\unitlength,page=1]{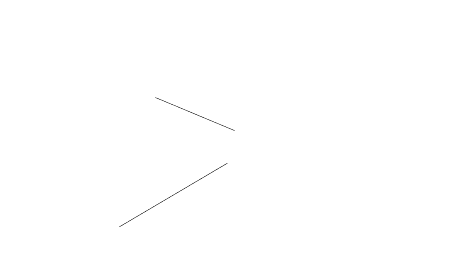}}%
    \put(0.93642007,0.25301182){\makebox(0,0)[lt]{\lineheight{1.25}\smash{\begin{tabular}[t]{l}$La_0$\end{tabular}}}}%
    \put(0,0){\includegraphics[width=\unitlength,page=2]{annotate_vison_bit.pdf}}%
  \end{picture}%
\endgroup%

    \caption{Zinc blende vison octupole in the cubic unit supercell, showing on the left the exploded view of one vison's local environment, and on the right the global vison arrangement, with electric Dirac strings drawn in yellow. The long (counter-orientated) string is exactly three times the length of each of the short strings.}
    \label{fig:vison_arrangement}
\end{figure}

We place positive and negative visons respectively on the void sublattice where their charge is preferred by the staggered chemical potential $g'$, making them metastable. 
It can be shown that a zinc blende superlattice (see Fig.~\ref{fig:vison_arrangement}), corresponding to the original diamond lattice rescaled by a factor of $4m+1$, $m\in\mathbb{Z}$, places the rescaled $A(B)$ sites coincident with the $A(B)$ sites of the original lattice. We exploit this to generate a sequence of increasingly dilute artificial vison supercrystals, from which we may analyse the scaling of the crystal cohesive energy (thus being able to differentiate between chemical potential and interaction effects).

We do not have access to a closed form for the vison creation operator, although a good approximation may in principle be obtained from the Green's function of the lattice Laplacian~\cite{hermelePyrochlorePhotonsSpin2004}. Here we adopt instead a different procedure to create the desired artificial vison arrangement: 
\begin{enumerate}
\item[VC0] We find a polarization-free arrangement of electric Dirac strings that introduces the desired vison structure.
\item[VC1] For every pair of connected visons, we simultaneously rotate the fluxes of all plaquettes crossed by the Dirac string until branch cuts are crossed at the beginning and end of the string. This places the system in an excited state corresponding to a superposition of the desired vison (super)crystal structure and some distribution of photons.
\item[VC2] We anneal the system from an initial temperature $T_{\text{hot}} > \rho_g$ down to $T_{\text{cold}} \ll \rho_g$, allowing the energy of the metastable configuration to be read off once the photons have thermalized.
\end{enumerate}
%
%

%
%

\subsection{Numerical results}
\begin{figure}
\includegraphics{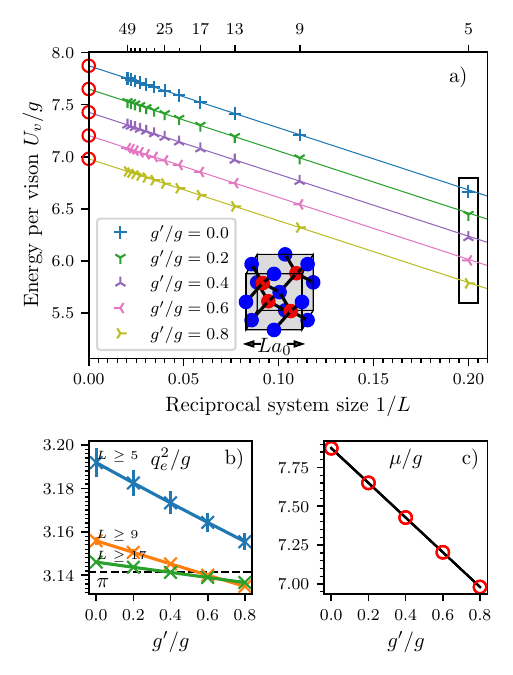}
\caption{Scaling of vison interaction energy with system size within the $U(1)_0$ phase. Panel a) shows the fits of the Coulomb functional form $U_v = M_{\mathrm{ZnS}} \, q_e^2 / (L a_0) + \mu$ to the data points with $L\ge 17$. This model marginally overestimates the cohesive energy at $L=5$ (see the data points outlined by a solid black line box). Note that $q_e$ may be interpreted as the elementary vison charge in Gaussian CGS units.
The inset shows the diamond superlattice arrangement of visons, see also Fig.~\ref{fig:vison_arrangement} for details.
The lower panels show the dependence of b) the Coulomb constant $q_e^2$ and c) the chemical potential $\mu$ on $g'$. Panel b) shows the dependence of the $q_e^2$ fits on the short-range interaction cutoff, suggesting convergence to the quadratic estimate $q_e^2 = \pi g$ (horizontal dashed black line). Panel c) shows the dependence of the vison chemical potential $\mu$ on $g'$, where a linear best fit $\mu(g,g')/g = -1.1154(2)g'/g + 7.873(1)$ appears to be in excellent agreement with the simulations. These results are independent of the cutoff choice.
For all plots, the absence of error bars indicates uncertainties smaller than the marker size (for reference, the standard error on the $\mu$ fits is of the order of $10^{-4}g$). 
}
\label{fig:coulomb_gp}
\end{figure}

\begin{figure}
\includegraphics{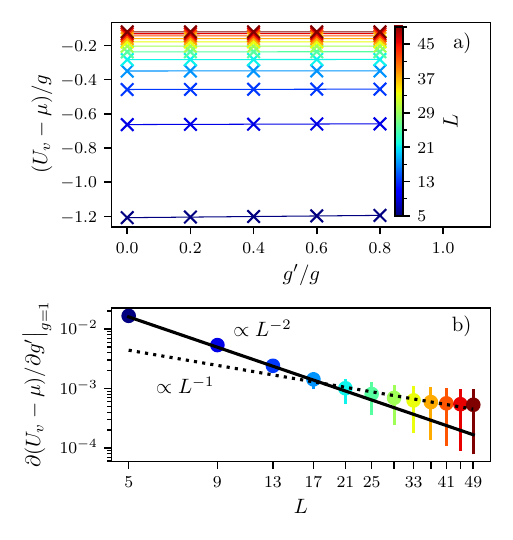}
\caption{Dependence of the vison interaction energy on $g'$. a) Scaling of the empirical vison interaction energy $U_v - \frac{\partial \mu}{\partial g}g - \frac{\partial \mu}{\partial g'}g'$ with $g'$. b) Log-log plot of the gradient of the interaction part of the vison energy with respect to $g'$ against linear system size $L$, exhibiting a potential crossover from $L^{-2}$ to $L^{-1}$ scaling. The error bars reflect $1\sigma$ standard error from linear regression.}
\label{fig:coulomb_gp_transposed}
\end{figure}

The scaling of the system energy with the linear dimension $L$ of the system is show in Figs.~\ref{fig:coulomb_gp} and~\ref{fig:coulomb_gp_transposed}. 

Inspired by the quadratic theory in Eq.~\eqref{eq:quadratic_guess}, we model the visons as a Coulombic plasma with cohesive energy
\begin{align}
    E &= \sum_{v < w} \frac{q_e^2(g,g') Z_v Z_w}{|r_v - r_w|} + N_{\rm vison} \mu(g,g')
    \, , 
\end{align}
where $Z_{v,w} = \pm 1$ is the sign of the charge of the visons labelled by $v,w$, $r_{v,w}$ are their positions and $N_{\rm vison}$ is the number of visons~\footnote{There are, in fact, two species of visons depending on whether they lie on the preferred sublattice. Here and throughout we refer only to the lower energy species.}. It follows immediately from trivial rescaling of the Hamiltonian that, at zero temperature, $q_e^2(g,0) \propto g$ and $\mu(g,0) \propto g$. The weak-field expansion in Eq.~\eqref{eq:quadratic_guess} suggests that $q_e^2$ is independent of $g'$ to leading order, with corrections arising from the polarized region in the immediate vicinity of a vison.

The numerical scaling of the vison energy with system size reveals that this intuition is essentially correct. The Coulomb functional form is an excellent fit to the system energy in Fig.~\ref{fig:coulomb_gp}a at all length scales, though the short-range energy for $L=5$ is marginally overestimated. Large values of the electric field are more heavily penalized by the quadratic theory than they are by the original cosine potential; near-neighbor vison dipoles are therefore more favorable in our numerics than the Coulomb estimate would predict. Indeed, we believe that this excess of vison dipoles is responsible for the homogeneous background seen in the static $e$ correlators of Fig.~\ref{fig:ssf}. 

The dependence of the zero-temperature vison chemical potential on $g'$, shown in Fig.~\ref{fig:coulomb_gp}c, is very robustly linear, suggesting that the chemical potentials associated with $g$ and $g'$ are essentially decoupled. The gradient is slightly less than the quadratic estimate ($-1$), reflecting the influence of nonlinearities in $\sin(e)$. 

Fig.~\ref{fig:coulomb_gp}b reveals the best-fit vison charge $q_e^2$ to be very close to the quadratic estimate at all values of $g'$. The apparent dependence of $q_e^2$ on $g'$ is due to the failure of the simple Coulomb model to properly capture the higher-order moments of the vison-vison interaction. We see that as short-ranged points are excluded from the fit, the dependence disappears. Fig.~\ref{fig:coulomb_gp_transposed} revisits this issue, subtracting off the robustly known chemical potential component to isolate the interaction part of the vison energy. We clearly see $L^{-2}$ scaling consistent with the coupling of local dipole moments to the inhomogeneous field in the immediate vicinity of a charge. Though the fit data points at larger radius appear to cross over to Coulomb $1/L$ scaling, this is an artefact of the Gaussian error model used for the linear regression. Perhaps more significantly, the $L^{-2}$ scaling remains consistent with all fits to within $1\sigma$ at all lengths. We can neither confirm nor rule out some Coulomb contribution arising from $g'$ (we leave such costly numerical exploration for future work). 


Combining this measurement of the vison charge with the photon dispersion measurement from Sec.~\ref{sec:uniform}, we are now in principle able to compute a large-$S$ approximation of the fine-structure constant. There is, however, a subtlety arising from the electric-magnetic duality: In the presence of both gapped magnetic and electric monopoles, there are a priori two dimensionless numbers of interest, the `electric fine structure constant' $\alpha_e = q_e^2/\hbar c$ and the `magnetic fine structure constant' $\alpha_m = q_m^2/\hbar c$, where $q_m$ is the charge of a spinon. Dirac's quantization condition imposes~\cite{stackMonopolesQuarkConfinement1992} a reciprocity relationship on the two fine structure constants, 
\begin{align}
\alpha_e\alpha_m = \frac{1}{4} 
\, . 
\label{eq:Dirac_quant}
\end{align}
This condition forces at least one of $\alpha_m$ and $\alpha_e$ to exceed the confining threshold $\alpha_c\simeq 0.21$~\cite{cardyUniversalPropertiesGauge1980, luckUniversalCriticalBehavior1982,cellaCoulombLawPure1997, stackMonopolesQuarkConfinement1992}, seemingly suggesting that either spinons, visons or both are confined. Some care should be taken before jumping to this conclusion, however. Though it is known that matter-free quantum $U(1)$ lattice gauge theory has a strong-weak duality mapping electric and magnetic fields to one another~\cite{montonenMagneticMonopolesGauge1977a,hermelePyrochlorePhotonsSpin2004,gorantlaModifiedVillainFormulation2021,wittenSDualityAbelianGauge1995}, it is a decidedly different question to ask whether this duality survives in the presence of dynamical matter.
Previous numerics have shown coexisting, Coulombic spinons and visons even with $\alpha_e \gg 1$~\cite{stackMonopolesQuarkConfinement1992}. All this is to say that we should not naively interpret $\alpha_e$ as a tuning parameter for vison confinement.

Substitution of the large-$S$ speed of light $\hbar c_{S=\infty} \simeq 0.15ga_0$~\cite{kwasigrochSemiclassicalApproachQuantum2017,bentonSeeingLightExperimental2012} yields $\alpha_m \simeq 21$. This number is independent of $g$ and $g'$, and well above $\alpha_c$. Nonetheless, the Coulomb form of the vison potential is characteristic of deconfined visons.

The Dirac quantization condition Eq.~\eqref{eq:Dirac_quant} allows us to calculate the value of $\alpha_m = 1/(4\alpha_e)$. At $S=\infty$, it is of order $\simeq 0.01$, interestingly close to the value obtained from recent exact diagonalization ($\alpha_{m,ED} = 0.07$~\cite{paceEmergentFineStructure2021}). Accounting for Hartree-Fock corrections to the speed of light~\cite{kwasigrochSemiclassicalApproachQuantum2017} renormalizes $\alpha_e$ to $\simeq 7.7$; in surprisingly reasonable agreement with $1/(4\alpha_{m,ED})\simeq 3.6$.
%
%

\section{Conclusions
\label{sec:conclusions}}
In this work, we studied the behaviour of (Kramers) quantum spin ice on a breathing pyrochlore lattice, by means of perturbative arguments and large-$S$ approximations. We demonstrated that the broken inversion symmetry allows for a new term ($ig' (\mathcal{O} - \mathcal{O}^\dagger)$) that would have otherwise been forbidden. As shown in Fig.~\ref{fig:parameter_control}, the anisotropy itself needs not be particularly large for $g'$ to be comparable with $g$. We therefore suggest that the $U(1)_{\pi/2}$ phase could be most easily realized in a `minimally breathing' pyrochlore compound -- the effect requires breaking inversion symmetry without decoupling the $A$ tetrahedra.

The $g'$ interaction effectively frustrates the emergent electric field, as there is no path in state space that can smoothly interpolate between the zero-flux state and the $\pi/2$-flux state. Instead, the fluxes jump abruptly between the four values allowed by the vison quantization condition. At finite temperature, the phases are separated by first-order liquid-gas like phase transition lines, terminating at critical end points which we tentatively placed in the 3D Ising universality class based on a finite-size scaling analysis.

When $g>|g'|$, we found that the interactions between visons are well approximated by an essentially $g'$-independent Coulomb law, in addition to a small correction term arising from deviations from the Coulomb law at short range. 
The more significant effect of $g'$ is the emergence of a staggered chemical potential for the visons. It follows from duality 
that the excitations above the $U(1)_{\pi/2}$ vacuum are similar -- one may readily view them as vison holes interacting with each other via a Coulomb potential with the roles of $g$ and $g'$ interchanged. 


The behaviour we have uncovered can be seen as the (quantum) electric analogue of spin fragmentation in classical spin ice~\cite{ dunQuantumClassicalSpin2020,lefrancoisSpinDecouplingStaggered2019,lefrancoisFragmentationSpinIce2017,brooks-bartlettMagneticMomentFragmentationMonopole2013}, occurring for instance in iridate pyrochlores, where the interpenetrating AIAO magnetic bias field causes a monopole crystal to condense while the spins retain a classical $U(1)$ fluctuating component. 
Similarly in our system, the $U(1)_{\pi/2}$ vison crystal phase remains a spin liquid dual to the original $U(1)_0$ phase, undergoing fragmented Coulomb phase dynamics. Unlike the magnetic sector analogue, the electric phenomenology is not due to an extrinsic (Ir magnetism) contribution but it is inherent to the quantum spin ice system and its interactions; moreover, in our case there is a microscopic one-to-one mapping between the effective Hamiltonians of both the $U(1)_0$ and $U(1)_{\pi/2}$ phases. 

We further studied the dynamical structure factor of the system, revealing, to leading order, a photon dispersion independent of $g'$. We interpret this as the long-wavelength photons decoupling from a deep-UV fluctuating background, just as real photons do in ionic crystals. More interestingly, we found a correction at order $e^3$ that directly influences the photon self-energy, modifying the experimentally accessible two-point $S^z$ correlator. This correction is in essence a QSI analogue of second harmonic generation in a nonlinear crystal. In the presence of a $g'e^3$ vertex, two photons may be up-converted to a single, higher-energy photon lying well above the original photon dispersion.

Realization of non-negligible $g'$ requires an admittedly fine-tuned energy hierarchy: $J_{zz,B}/J_{zz,A}$ must not be so small as to decouple the tetrahedra, 
yet also not so close to $1$ that inversion symmetry suppresses $g'$. Nonetheless, the apparent ability of applied external electric fields to change the sign of the effective value of $g$~\cite{mandalElectricFieldResponse2019,lantagne-hurtubiseElectricFieldControl2017} suggests that in the presence of non-zero $J_{z\pm}$, an electric field tuning $g$ between $U(1)_0$ and $U(1)_\pi$ could drive the system into an intermediate $U(1)_{\pi/2}$ phase.

In our work, we have identified a number of distinctive phenomena that could be used as possible experimental probes to test the effects of $g'$ in candidate materials. To maximize the chances of hitting a `sweet spot' in parameter space, we suggest that a candidate material should be subjected to an applied [100] electric field, which is known to effectively tune the $g$ parameter~\cite{mandalElectricFieldResponse2019}. We propose that an experiment sweeping $g$ from negative to positive should search for three phenomena:
a) two divergences in thermodynamic observables as the system transits across the phases $U(1)_0 \to U(1)_{\pi/2} \to U(1)_\pi$;
b) the appearance of a structured high-energy continuum in $\langle S^zS^z\rangle$ correlations (see Figs.~\ref{fig:dsf_all} and~\ref{fig:upconversion}), which can be accessed, e.g., using inelastic neutron scattering;
and 
c) rod-like correlations from domain wall formation in $\langle e e \rangle$ at criticality (see Fig.~\ref{fig:ssf}). 
Though these 12-spin correlators are for all practical purposes inaccessible, we suspect that some vestige thereof may yet appear in the $\langle S^xS^x\rangle$ and $\langle S^yS^y\rangle$ channels.

This work highlights the richness of emergent QED in frustrated magnets with broken inversion symmetry, representing an unconventional lever with which the strong-coupling QED of QSI may be tuned towards yet new phases and phenomena. 
%
%
%
%
%
%
%
%
%

%
%

\section*{Acknowledgements}
The authors gratefully acknowledge fruitful discussions with Peter Holdsworth, Michal Kwasigroch, Christopher Laumann, Salvatore Pace and Attila Szab\'{o}. We are particularly indebted to Attila Szab\'{o} for providing the code that formed the basis of the numerical work here. 
For the purpose of open access, the authors have applied a Creative Commons Attribution (CC BY) licence to any Author Accepted Manuscript version arising from this submission. 
This research was funded in part by the Engineering and Physical Sciences (EPSRC) grants No. EP/P034616/1, EP/V062654/1 and EP/T028580/1. 
%
%

\bibliography{u1sl_bibertool}
%
%
\appendix

\section{Geometry}
\label{app:geometry}
It is conventional to write the QSI Hamiltonian with respect to the local reference frame for the spins. The convention adopted here is the same as in Ref.~\onlinecite{szaboSeeingLightVison2019}: $\boldsymbol{S}_j = \sum_{\alpha={x,y,z}}S^\alpha_j \mathbf{e}_{\mu(j),\alpha}$, where $\mu(j)$ denotes the pyrochlore sublattice of site $j$ and 
\begin{equation*}
\begin{array}{lll}
\mathbf{e}_{0, x} = 
\frac{1}{\sqrt{6}}\begin{pmatrix} 1 \\ 1\\ -2 \end{pmatrix} 
,  &
\mathbf{e}_{0, y} = 
\frac{1}{\sqrt{2}}\begin{pmatrix} -1\\ 1 \\0 \end{pmatrix}
, &
\mathbf{e}_{0, z} = 
\frac{1}{\sqrt{3}}\begin{pmatrix} 1 \\1\\ 1 \end{pmatrix} 
\\
\mathbf{e}_{1, x} = 
\frac{1}{\sqrt{6}}\begin{pmatrix} 1 \\ -1\\ 2 \end{pmatrix}
, &
\mathbf{e}_{1, y} = 
\frac{1}{\sqrt{2}}\begin{pmatrix} -1\\ -1 \\0 \end{pmatrix}
, &
\mathbf{e}_{1, z} = 
\frac{1}{\sqrt{3}}\begin{pmatrix} 1 \\-1\\ -1 \end{pmatrix} 
\\
\mathbf{e}_{2, x} = 
\frac{1}{\sqrt{6}}\begin{pmatrix} -1 \\ 1\\ 2 \end{pmatrix}
, &
\mathbf{e}_{2, y} = 
\frac{1}{\sqrt{2}}\begin{pmatrix} 1\\ 1 \\0 \end{pmatrix}
, &
\mathbf{e}_{2, z} = 
\frac{1}{\sqrt{3}}\begin{pmatrix} -1 \\1\\ -1 \end{pmatrix} 
\\
\mathbf{e}_{3, x} = 
\frac{1}{\sqrt{6}}\begin{pmatrix} -1 \\ -1\\ -2 \end{pmatrix}
, &
\mathbf{e}_{3, y} = 
\frac{1}{\sqrt{2}}\begin{pmatrix} 1\\ -1 \\0 \end{pmatrix}
, &
\mathbf{e}_{3, z} = 
\frac{1}{\sqrt{3}}\begin{pmatrix} -1 \\-1\\ 1 \end{pmatrix} 
. 
\end{array}
\end{equation*}
Further, we introduce the four inequivalent vectors 
\begin{align*}
    \mathbf{b}_0 &= \frac{a_0}{8}(+1,+1,+1)\\
    \mathbf{b}_1 &= \frac{a_0}{8}(+1,-1,-1)\\
    \mathbf{b}_2 &= \frac{a_0}{8}(-1,+1,-1)\\
    \mathbf{b}_3 &= \frac{a_0}{8}(+1,-1,-1)
\end{align*}
connecting the centre of a tetrahedron to the four nearest-neighbor pyrochlore sites at its corners. 
%
%

\section{Derivation of the Complex ring flip
\label{app:details}}
This section presents the finer details involved in obtaining the quoted values for $g$ and $g'$ in perturbation theory. Many of the more verbose calculations involved were performed using our noncommutative algebra package
\texttt{commutation}~\footnote{Available from PyPI, \texttt{pip install commutation}.}.

For completeness, we include an enumeration of the symmetry properties of the three allowed crystal field doublets in psuedospin-$\frac{1}{2}$ pyrochlores in Table~\ref{tab:symmetry_cases}. 
\begin{table}[h!]
\begin{tabular}{l|l|l|c|c}
	Doublet & $\qquad\mathcal{T}$ & $\qquad\quad C_3$ &
	$-\zeta_{ij}^* \equiv \gamma_{ij}$  & $J_{z\pm}$\\
	\hline
	Eff. Spin-$\frac{1}{2}$ & 
		$\bm{S} \mapsto - \bm{S}$ &
		$ \begin{aligned} 
			S^z &\mapsto S^z \\
			S^\pm &\mapsto e^{\pm\frac{2\pi i}{3} } S^\pm
		\end{aligned}$  &
		$ 1, w ,w^2 $ &
		$\neq 0$
		\\ 
\hline
	Dip.-Octupolar & 
		$\bm{S} \mapsto - \bm{S}$ &
		$\bm{S} \mapsto +\bm{S}$  &
		$ 1 $ &
		$\neq 0$
		\\
\hline
	Non-Kramers & 
		$\begin{aligned}
			S^z &\mapsto - S^z\\
			S^\pm &\mapsto S^\mp
		\end{aligned}$ &
		$ \begin{aligned} S^z &\mapsto S^z \\
			S^\pm &\mapsto e^{\pm\frac{2\pi i}{3} } S^\pm \end{aligned}$  &
		$ 1, w ,w^2 $ &
		$= 0$
		\\ 
\end{tabular}
\caption{Known crystal field doublets in rare-earth pyrochlores under the action of time reversal $\mathcal{T}$ and rotation $C_3$ about the local $z$ axis. 
Rotation $C_2$ about the local $x$ axis gives $ S^z \mapsto -S^z$ and $S^\pm \mapsto S^\mp$ for all cases. Here $w=e^{2\pi i/3}$; and $\zeta_{ij}$, $\gamma_{ij}$, and $J_{z\pm}$ are defined in Eq.~\eqref{eq:Hamiltonian_full}. Table adapted from Ref.~\onlinecite{rauFrustratedQuantumRareearth2019}.}
\label{tab:symmetry_cases}
\end{table}
It follows that complex ring flips are not possible in dipolar-octupolar QSI: a basis exists in which all coupling constants are real. 
%
%

\subsection{Geometric language for perturbation theory}
\label{ssec:diagrams}
\begin{figure}
\centering
\includegraphics{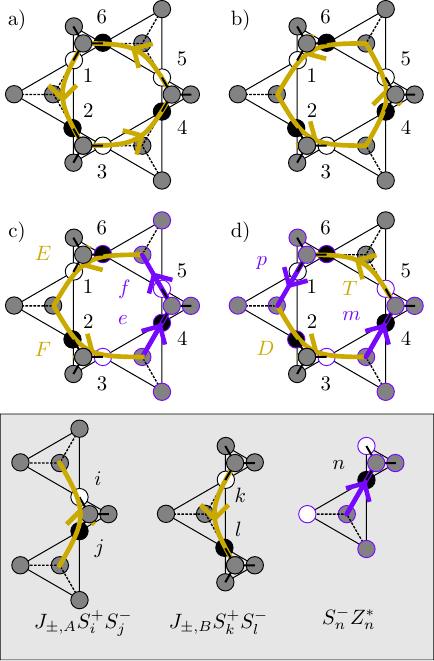}
\caption{Elementary perturbative processes in the spinon transport language. These diagrams correspond to the following operator sets:
\begin{tabular}{ll}
a) &  $\{J_{\pm, B}S^+_1S^-_2, J_{\pm, B}S^+_3S^-_4, J_{\pm, B}S^+_5S^-_6\}$\\
b) & $\{J_{\pm,A}S^+_1S^-_6, J_{\pm,A}S^+_5S^-_4, J_{\pm,A}S^+_3S^-_2\}$\\
c) & $\{J_{\pm,A}S^+_1 S^-_6, J_{\pm,A}S^+_3S^-_2, S^-_4 Z^*_4, S^+_5 Z_5\} =: \{E,F,e,f\}$ \\
d) &  $\{J_{\pm,B}S^+_5 S^-_6, J_{\pm,A}S^+_3S^-_2, S^+_1 Z_1, S^-_4 Z^*_4\} =: \{T,D,p,m\}$. 
\end{tabular}
a),b) are the standard third-order XXZ ring flips; c),d) are the novel complex ring flips.
We have defined $Z_j = \sum_{\langle ij \rangle} J_{z\pm, A/B} \zeta_{ij} S^z_i$ as the sum over all six nearest neighbors of site $j$, weighted by the geometric $\zeta$ factors and the appropriate $J_{z\pm, A/B}$ for the tetrahedron that bond $\langle ij \rangle$ belongs to.}
\label{fig:channels_explicit}
\end{figure}

The task before us is to simplify terms in the series expansion of $H_{eff}$~\eqref{eq:heff}. As the calculation is fairly technical, it is useful to develop a graphical language to express terms in the perturbative expansion. To this end, we establish a rigorous correspondence between terms in the perturbation expansion and links (paths) on the diamond lattice:
%
%
%
\begin{description}
\item[$J_{\pm}S^+_iS^-_j$] Spinon transport between second neighbor (i.e., same sublattice) diamond sites. We associate this term with the path across links $i$ and $j$ highlighted by a yellow arrow in Fig.~\ref{fig:channels_explicit}.
\item[$J_{z\pm} S^z_i S^+_j$] Spinon transport between nearest-neighbor (i.e., different sublattice) diamond sites, with a $S^z_i$ dependent $\mathbb{Z}_2$ phase. These are associated with a diamond link $j$ highlighted a purple arrow in Fig.~\ref{fig:channels_explicit}.
\item[$J_{\pm\pm}S^+_iS^+_j$] Double-hop of two $A$ site spinons to the same $B$ site (or vice versa), illustrated for example by the green arrows in Fig.~\ref{fig:hpmpm}. 
\end{description}
These associations are illustrated in the legend of Fig.~\ref{fig:channels_explicit}. 
%
%

\subsection{Algorithm for generating terms}
We generate terms in the effective Hamiltonian as follows:
\begin{enumerate}
    \item Fix an order $N$ in perturbation theory. This sets the number of diamond lattice paths available for use.
    \item Attempt to arrange the paths into a closed loop, guaranteeing the annihilation of all spinons. (Up to reasonable orders, this process can be done exhaustively.)
    \item Write down the $N!$ perturbations of the set associated with a particular spinon diagram. Calculate the virtual spinon energy for each factor of the propagator $[-H_0]^{-1}$. 
    \item Sum over all permutations of the operators.
\end{enumerate}

As a check, we rederive the standard third-order ring flip in breathing XXZ quantum spin ice as follows: 
%
\begin{enumerate}
    \item Fix $N=3$.
    \item The only way to connect three two-link arrows and annihilate any virtual spinon created in the process is to arrange them around a hexagonal plaquette. There are two ways to do this: loop a) in Fig.~\ref{fig:channels_explicit} has its virtual spinons hop only on the $A$ sublattice; loop b) in the same figure transports only via the $B$ sublattice.
    \item For both loop a) and loop b), all terms in their corresponding operator sets commute. All virtual states in process a) have the same energy, $J_{zz,A}$; similarly all virtual states of b) have energy $J_{zz,B}$.
    \item Sum to give~\footnote{Note that this differs from Ref.~\onlinecite{hermelePyrochlorePhotonsSpin2004} by a factor of $8$ due to a trivial factor of $2$ difference in the definition of $J_\pm$.}
    $$\left[3! \frac{J_{\pm, B}^3}{(J_{zz,A})^2} + 3! \frac{J_{\pm, A}^3}{(J_{zz,B})^2}\right] S_1^+S_2^-S_3^+S_4^-S_5^+S_6^- \, .$$ 
\end{enumerate}
%
%

\subsection{Teardrop exchange}
The spinon energy calculation is more complicated at fourth order and above. Consider as a first example the teardrop exchange illustrated in Fig.~\ref{fig:terms}c. Note that this is one of $12$ distinct teardrops associated with the plaquette $p$ made up of links $123456$, and that there is a spectator spin that is flipped twice, returning it to its original state. The loop depicted in the figure corresponds to the operator set $\{J_{\pm,B} S_1^+S_7^-, J_{\pm,B} S_7^-S_2^-, J_{\pm,B} S_3^+S_4^-, J_{\pm,B} S_5^+S_6^-\}$. Of the $4!$ orderings of this set, eight enter the four-spinon sector, introducing a factor of two in the denominator. The remaining $16$, like the third order calculation, correspond to creating spinon pairs, transporting them around a loop, and annihilating them. This process then gives an overall factor of
$$
\left[\frac{4}{2J_{zz,A}^3} + \frac{16}{J_{zz,A}^3}\right] J_{\pm, B}^4 \left[S_7^-S_7^+ + S_7^+S_7^-\right] \mathcal{O}_p
$$
in the series expansion of $H_{eff}$. Summing over the twelve teardrops on plaquette $p$ and simplifying $S_7^-S_7^+ + S_7^+S_7^-] = 1$ ultimately gives the contribution
$$
12 \left[ \frac{18}{J_{zz,A}^3} J_{\pm, B}^4 + \frac{18}{J_{zz,B}^3} J_{\pm, A}^4 \right]\mathcal{O}_p 
\, . 
$$
%
%
%

\subsection{Complex ring flips}
We identify two distinct channels that contribute to the fourth-order complex ring flips, depending on the placement of the two $J_{z\pm}$ $S^+$ terms on the hexagon, which we label \textit{ortho} and \textit{para} (see Fig.~\ref{fig:channels_explicit}c,d). For the purposes of this section, we define $\mathcal{A} := J_{zz,A}/2, \mathcal{B} := J_{zz,B}/2$.
%
%

\subsubsection{\textit{ortho} channel}
Define the following operators, corresponding to those shown in Fig.~\ref{fig:channels_explicit}: 
\begin{align*}
    E &= J_{\pm, A} S_6^-S_1^+\\
    F &= J_{\pm, A} S_2^- S_3^+\\
    e &= S_4^- Z_4^*\\
    f &= S_5^+ Z_5 
\, . 
\end{align*}
One can then see that permutations involving $Ee$ or $Ff$ acting on a ground state create 4-spinon virtual states, while all other arrangements remain in the 2-spinon sector:
%

\vspace{0.2 cm}
\!\!\!\!\!\!\!\!\!\begin{tabular}{|l|l|}
\hline
 Spinon energy product & Terms \\
\hline
    $(\mathcal{A}+\mathcal{B})^2(\mathcal{A}+3\mathcal{B})$ & $ fFEe + eEFf$\\
    \hline
     $(2\mathcal{B})^2(\mathcal{\mathcal{A}}+3\mathcal{B})$& $ FfeE + EefF$\\
    \hline
     $2\mathcal{B}(\mathcal{A}+\mathcal{B})(\mathcal{A}+3\mathcal{B})$&$FfEe+ EeFf+ fFeE+ eEfF$\\
    \hline
     $(\mathcal{A}+\mathcal{B})^3$ & $fEFe + eFEf$\\
    \hline
     $2\mathcal{B}(\mathcal{A}+\mathcal{B})^2$ & $eFfE + EfFe + fEeF + FeEf$\\
    \hline
     $(2\mathcal{B})^2(\mathcal{A}+\mathcal{B})$ & $ efEF + efFE + feEF + feFE$\\
    &$+ EFef + FEef + EFfe + FEfe$\\
    \hline
    $(2\mathcal{B})^2(\mathcal{A}+\mathcal{B})$&$EfeF + FefE$\\
\hline
\end{tabular}
\vspace{0.1 cm}

After the $H_0$ values are taken care of, commutation properties can be used to simplify the expression for $H_{\rm eff}$ to
\begin{align}
	H_{\rm ortho} =& \alpha_{1} \left(fEFe + eFEf\right) + \alpha_{2} \{\{e,f\},EF\} \nonumber \\
 &+ \alpha_{3} \left(EfeF+FefE\right) \, , \\ 
    \alpha_{1} =& \frac{1}{\mathcal{A}+3\mathcal{B}}\left(\frac{1}{(\mathcal{A}+\mathcal{B})^2} + \frac{1}{(2\mathcal{B})^2}+\frac{2}{2\mathcal{B}(\mathcal{A}+\mathcal{B})}\right) \nonumber \\&+ \frac{1}{(\mathcal{A}+\mathcal{B})^3} \nonumber \\
    \alpha_{2} =& \frac{1}{2\mathcal{B}(\mathcal{A}+\mathcal{B})^2} + \frac{2}{(2\mathcal{B})^2(\mathcal{A}+\mathcal{B})} \nonumber \\
    \alpha_{3} =& \frac{1}{(2\mathcal{B})^2(\mathcal{A}+\mathcal{B})} 
\, . 
\nonumber
\end{align}
%
%
%

\subsubsection{\textit{para} channel}
Let us then consider the \textit{para} ring flips. Using the labeling convention in Fig.~\ref{fig:channels_explicit}, the operators are
\begin{align}
	p & = Z_1S_1^+\\
	m &= Z_4^*S_4^- \\
	T &= J_{\pm, A} S^+_5S^-_6 \\
	D &= J_{\pm, B} S^-_2 S^+_3
\, . 
\end{align}
Correspondingly, the six possible inequivalent spinon channels are

\vspace{0.1 cm}
\!\!\!
\begin{tabular}{|l|l|l|}
\hline
    Spinon en. prod. & Terms \\
\hline
    $(\mathcal{A}+\mathcal{B})^3$   & $pTDm + pDTm + mTDp + mDTp$\\
    $2\mathcal{B}(\mathcal{A}+\mathcal{B})^2$ & $pTmD+mTpD + DmTp + DpTm$\\
    $2\mathcal{A}(\mathcal{A}+\mathcal{B})^2$ & $pDmT+mDpT + TmDp + TpDm$\\
    $4\mathcal{B}(\mathcal{A}+\mathcal{B})^2$ & $pmTD+mpTD+DTpm+DTmp$\\
    $4\mathcal{A}(\mathcal{A}+\mathcal{B})^2$ & $pmDT+mpDT+TDpm+TDmp$\\
    $4\mathcal{A}\mathcal{B}(\mathcal{A}+\mathcal{B})$  & $TpmD+TmpD+DpmT+DmpT$\\
\hline
\end{tabular}
\vspace{0.1 cm}

Observing that the commutators $[D,T]$ and $[m,p]$ vanish, we obtain: 
\begin{align}
	H_{p,\rm para} =& \alpha_{4} \left(pDTm + mTDp\right) \nonumber\\
    &+ \alpha_{5} \big[ \mathcal{A} \left(pTmD+mTpD + DmTp + DpTm\right)\nonumber\\
	&\ +\mathcal{B}\left(pDmT+mDpT + TmDp + TpDm\right) \nonumber\\
    &+ \frac{\mathcal{A}+\mathcal{B}}{2}\left(pmDT + TDmp\right) \nonumber\\
    &+ (\mathcal{A}+\mathcal{B})(TpmD+DmpT) \big] 
    \, , 
    \\
    \alpha_{4} =& \frac{2}{(\mathcal{A}+\mathcal{B})^3} \nonumber\\
    \alpha_{5} =& \frac{1}{2\mathcal{A}\mathcal{B}(\mathcal{A}+\mathcal{B})^2} 
\, . 
\nonumber
\end{align}

The final expression for $g,g'$ and the decorated ring flip terms is quite cumbersome, and does not provide much insight beyond the symmetry properties discussed in the main text. We give these expressions below for completeness in terms of the symmetrized parameters: 
\begin{align*}
	\overline{J_{zz}} &= \frac{J_{zz,A} + J_{zz,B}}{2} &&
	\Delta = \frac{J_{zz,A} - J_{zz,B}}{2} \\
	\overline{J_{\pm}} &= \frac{J_{\pm A} + J_{\pm B}}{2} &&
	\delta = \frac{J_{\pm A} - J_{\pm B}}{2}\\	
	\overline{J_{z\pm}} &= \frac{J_{z\pm, A} + J_{z\pm,  B}}{2}&&
	\kappa = \frac{J_{z\pm, A} - J_{z\pm, B}}{2}
\, , 
\end{align*}
\begin{widetext}
\begin{align*}
g +ig' =& 
\frac{1}{(\overline{J}_{zz} - \Delta)^3 (\overline{J}_{zz} + \Delta)^3 8 \overline{J}_{zz}^3} \Bigg[
1728 J^4 (\overline{J}_{zz}^6 + 3 \overline{J}_{zz}^4 \Delta^2) 
+96 J^3 (\overline{J}_{zz}^7 + 216 \overline{J}_{zz}^5 \delta \Delta +    \overline{J}_{zz}^3 (72 \delta - \Delta) \Delta^3\\
+& 12 J^2 \Big\{
            \overline{J}_{zz}^6 (-31 \overline{J}_{z\pm}^2 +       48 \delta (18 \delta + \Delta) +       6 i \sqrt{3} \overline{J}_{z\pm} \kappa - 35 \kappa^2) +    \overline{J}_{zz}^2 \Delta^4 (\overline{J}_{z\pm}^2 + 14 i \sqrt{3} \overline{J}_{z\pm} \kappa -       3 \kappa^2) + 6 \Delta^6 (\overline{J}_{z\pm}^2 + \kappa^2) \\
        &+    4 \overline{J}_{zz}^4 \Delta^2 (6 \overline{J}_{z\pm}^2 + 648 \delta^2 -       12 \delta \Delta - 5 i \sqrt{3} \overline{J}_{z\pm} \kappa +       8 \kappa^2) +    2 \overline{J}_{zz}^3 \Delta^3 (3 i \sqrt{3} \overline{J}_{z\pm}^2 + 94 \overline{J}_{z\pm} \kappa +       7 i \sqrt{3} \kappa^2) \\
        &-    i \overline{J}_{zz}^5 \Delta (\sqrt{3} \overline{J}_{z\pm}^2 - 146 i \overline{J}_{z\pm} \kappa +       5 \sqrt{3} \kappa^2) -    i \overline{J}_{zz} \Delta^5 (5 \sqrt{3} \overline{J}_{z\pm}^2 - 42 i \overline{J}_{z\pm} \kappa +       9 \sqrt{3} \kappa^2)\Big\} \\
+& 24 J \delta\Big\{
-\overline{J}_{zz} \Delta^5 (13 \overline{J}_{z\pm}^2 +       2 i \sqrt{3} \overline{J}_{z\pm} \kappa + 9 \kappa^2) +    2 \overline{J}_{zz}^3 \Delta^3 (35 \overline{J}_{z\pm}^2 + 144 \delta^2 +       6 i \sqrt{3} \overline{J}_{z\pm} \kappa + 31 \kappa^2) -    \overline{J}_{zz}^5 \Delta (57 \overline{J}_{z\pm}^2 \\
 &- 864 \delta^2 +       10 i \sqrt{3} \overline{J}_{z\pm} \kappa + 53 \kappa^2) +    2 \overline{J}_{zz}^6 (3 i \sqrt{3} \overline{J}_{z\pm}^2 + 6 \overline{J}_{zz} \delta - 42 \overline{J}_{z\pm} \kappa -       i \sqrt{3} \kappa^2) \\
 &+    4 \overline{J}_{zz}^2 \Delta^4 (i \sqrt{3} \overline{J}_{z\pm}^2 - 3 \overline{J}_{zz} \delta +       2 \overline{J}_{z\pm} \kappa -       3 i \sqrt{3} \kappa^2) + \Delta^6 (-i \sqrt{3} \overline{J}_{z\pm}^2 +       6 \overline{J}_{z\pm} \kappa + 3 i \sqrt{3} \kappa^2) \\
 &+    \overline{J}_{zz}^4 \Delta^2 (-9 i \sqrt{3} \overline{J}_{z\pm}^2 + 70 \overline{J}_{z\pm} \kappa +       11 i \sqrt{3} \kappa^2) \Big\}
    \\
    +& 12 \delta^2 \Big\{
4 \overline{J}_{z\pm} \Delta^6 (2 \overline{J}_{z\pm} - i \sqrt{3} \kappa) +    \overline{J}_{zz}^6 (-53 \overline{J}_{z\pm}^2 + 16 \delta (9 \delta + \Delta) -       14 i \sqrt{3} \overline{J}_{z\pm} \kappa - 33 \kappa^2) \\
     &+    \overline{J}_{zz}^2 \Delta^4 (-25 \overline{J}_{z\pm}^2 + 2 i \sqrt{3} \overline{J}_{z\pm} \kappa +       11 \kappa^2) +    2 \overline{J}_{zz}^4 \Delta^2 (35 \overline{J}_{z\pm}^2 + 216 \delta^2 -       8 \delta \Delta + 8 i \sqrt{3} \overline{J}_{z\pm} \kappa +       11 \kappa^2) \\
     &+ 2 \overline{J}_{zz}^3 \Delta^3 (-13 i \sqrt{3} \overline{J}_{z\pm}^2 + 62 \overline{J}_{z\pm} \kappa -       9 i \sqrt{3} \kappa^2) +    i \overline{J}_{zz} \Delta^5 (11 \sqrt{3} \overline{J}_{z\pm}^2 + 10 i \overline{J}_{z\pm} \kappa +       7 \sqrt{3} \kappa^2) \\
     &+ i \overline{J}_{zz}^5 \Delta (15 \sqrt{3} \overline{J}_{z\pm}^2 + 114 i \overline{J}_{z\pm} \kappa +       11 \sqrt{3} \kappa^2)
\Big\}
\Bigg]
\end{align*}
\begin{align*}
\tilde{\xi}_{1}=& 
  \frac{24 (k^2 - \kappa^2)}{16 \overline{J}_{zz}^3 (\overline{J}_{zz}^2 - \Delta^2)^2}i \Big(2 (i + \sqrt{3}) (\overline{J}_{\pm} + \delta)^2 (\overline{J}_{zz} + \Delta)^2 (10 \overline{J}_{zz}^2 - 5 \overline{J}_{zz} \Delta + \Delta^2) \\
  &- 2 (-i + \sqrt{3}) (\overline{J}_{\pm} - \delta)^2 (\overline{J}_{zz} - \Delta)^2 (10 \overline{J}_{zz}^2 + 5 \overline{J}_{zz} \Delta + \Delta^2) \Big)
\end{align*}
\begin{align*}
\tilde{\xi}_{2A}=& \frac{1}{4 \overline{J}_{zz}^3 (\overline{J}_{zz} - \Delta) (\overline{J}_{zz} + \Delta)^2}
\Big(3 (\overline{J}_{\pm} - \delta) ((4 + 4 i \sqrt{3}) (\overline{J}_{\pm} + \delta) (\overline{J}_{zz} + \Delta) (8 \overline{J}_{zz}^2 - \Delta^2) \\
&+ (4 - 4 i \sqrt{3}) (\overline{J}_{\pm} - \delta) (\overline{J}_{zz} - \Delta) (10 \overline{J}_{zz}^2 + 5 \overline{J}_{zz} \Delta + \Delta^2)) (\overline{J}_{z\pm}+ \kappa)^2\Big)
\end{align*}
\begin{align*}
\tilde{\xi}_{3} =& \frac{12 (\overline{J}_{\pm}^2 - \delta^2) (8 \overline{J}_{zz}^2 - \Delta^2) (\overline{J}_{z\pm}^2 - \kappa^2) }{\overline{J}_{zz}^3 (\overline{J}_{zz}^2 - \Delta^2) }
\, . 
 \end{align*}
\end{widetext}

The term $\tilde{\xi}_{2B}$ may be obtained from interchanging $A$ and $B$ in $\xi_{2A}$.
%
%

\subsection{Other terms}
\begin{figure}
\begin{tikzpicture}
    \draw (0, 4) node[inner sep=0, below right] {\includegraphics[width=0.7\columnwidth]{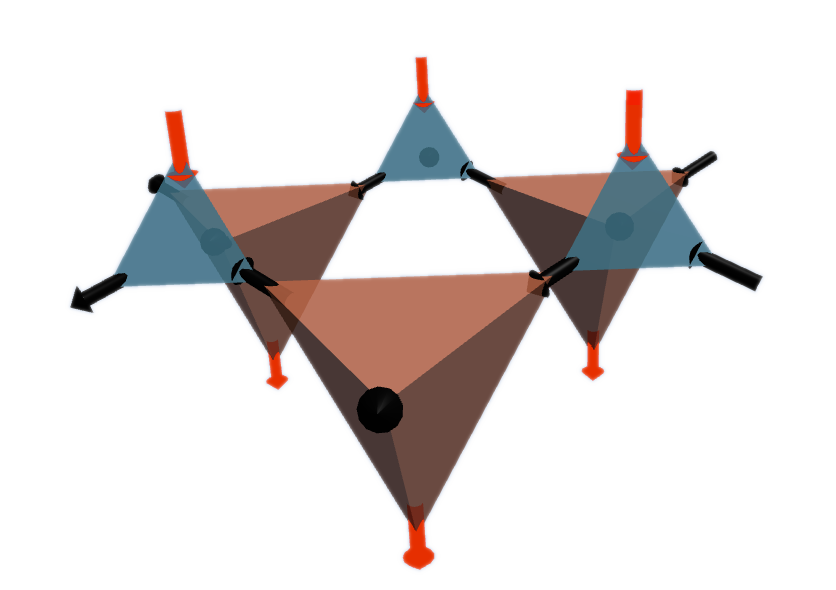}};
    \draw (0.2, 4-0.4) node[below right] {a)};
    \draw (0.2, -0.4) node[below right] {b)};
    \pgfmathsetmacro{\r}{0.05};
    \pgfmathsetmacro{\la}{0.4};
    \pgfmathsetmacro{\lb}{-0.2};
    \newcommand{\makeTet}[4]{
    \filldraw[thin, densely dotted, green]
        (#1,#2) -- ++({180+#4}:#3) circle(\r);
    \filldraw[thin, densely dotted]
        (#1,#2) -- ++ ({60+#4}:#3) circle(\r)
        (#1,#2) -- ++ ({-60+#4}:#3) circle(\r);
    \draw[thin, densely dotted] ({#1 - #3} ,#2) -- ++({30}:{#3*sqrt(3)}) -- ++(0,{-#3*sqrt(3)}) -- cycle;
    \filldraw[red] (#1,#2) circle(\r);
    }
    \newcommand{\makehexa}[2]{
    \foreach \theta in {0, 120, 240}
    {
        \pgfmathsetmacro{\Xa}{#1};
        \pgfmathsetmacro{\Ya}{#2};
        \pgfmathsetmacro{\Xa}{\Xa - (\la - \lb)*cos(\theta) };
        \pgfmathsetmacro{\Ya}{\Ya - (\la - \lb)*sin(\theta) };
        \makeTet{\Xa}{\Ya}{\la}{\theta}
        \pgfmathsetmacro{\Xa}{\Xa - (\la - \lb)*cos(120+\theta) };
        \pgfmathsetmacro{\Ya}{\Ya - (\la - \lb)*sin(120+\theta) };
        \makeTet{\Xa}{\Ya}{\lb}{-120+\theta}
    }
    }
    \makehexa{1}{-2}
    \draw [very thick, teal] ({1 - (\la - \lb)},-2) -- ++(60:{\la-\lb}) -- ++(0:{\la-\lb}) -- ++(-60:{\la-\lb}) -- ++(-120:{\la-\lb}) -- ++(180:{\la-\lb}) -- ++(120:{\la-\lb});
    \draw[teal] (1,-2) node {$\tilde \xi_1$};
    \makehexa{3}{-2}
    \draw [very thick, purple, dashed] ({3 - (\la - \lb)},-2) -- ++(30:{sqrt(3)*(\la-\lb)})
    -- ++(-90:{sqrt(3)*(\la-\lb)}) node [right] {$\tilde \xi_{2B}$} -- cycle ;
    \draw [very thick, purple] ({3 + (\la - \lb)},-2) -- ++(30:-{sqrt(3)*(\la-\lb)})
    node [left]{$\tilde \xi_{2A}$}
    -- ++(-90:-{sqrt(3)*(\la-\lb)})  -- ++(150:-{sqrt(3)*(\la-\lb)});
    \makehexa{5}{-2}
    \draw [very thick, orange] 
    ({5 - (\la - \lb)*cos(0)},{-2 - (\la - \lb)*sin(0)})
    --({5 + (\la - \lb)*cos(0)},{-2 + (\la - \lb)*sin(0)}) node [below right]{$\tilde \xi_3$}
    ({5 - (\la - \lb)*cos(60)},{-2 - (\la - \lb)*sin(60)})
    --({5 + (\la - \lb)*cos(60)},{-2 + (\la - \lb)*sin(60)})
    ({5 - (\la - \lb)*cos(-60)},{-2 - (\la - \lb)*sin(-60)})
    --({5 + (\la - \lb)*cos(-60)},{-2 + (\la - \lb)*sin(-60)});
\end{tikzpicture}
\caption{a) Normal $S^z$ operators used in the definitions of the ring flip terms appearing
in Eq.~\eqref{eq:compelx_ring flip}. b) A pyrochlore plaquette (black dots) and its nearest-neighbor off-ring sites (red and green dots), showing the four symmetry-inequivalent spin pairings. Red nearest neighbours correspond to those shown in panel a). The locations of the $S^z$ sites of the four effective interactions $\tilde{\xi}_1$, $\tilde{\xi}_{2A,B}$, and $\tilde{\xi}_3$ are also indicated in panel b)}\label{fig:normalspins}
\end{figure}
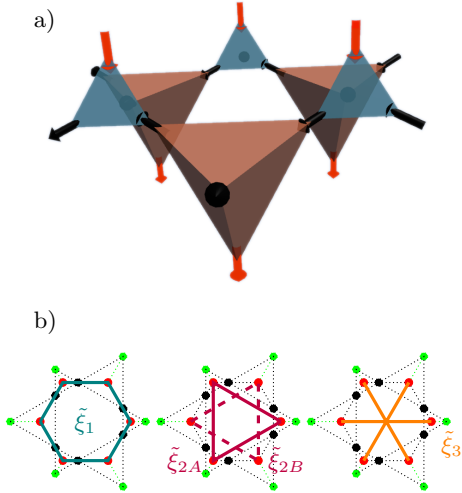

In addition to the real and complex ring flips $(g+ig')\mathcal{O}_p$ discussed in the main text, perturbation theory generates eight-spin operators of the form $S^zS^z\mathcal{O}_p$ coupling the ring flip to dangling $S^z$ moments nearest neighbor to the sites visited by the ring flip, parameterized by four independent complex couplings $\tilde{\xi}_1$, $\tilde{\xi}_{2A}$, $\tilde{\xi}_{2B}$, and $\tilde{\xi}_3$ as illustrated in Fig.~\ref{fig:normalspins}:
\begin{align}
\sum_{p} \left\{ \sum_{  (ij) \in \{1, 2A, 2B, 3\} } (\tilde{\xi}_{(ij)}) S^z_{ i} S^z_{j} \right\} \mathcal{O}_{p} + h.c. 
\, , 
\label{eq:compelx_ring flip}
\end{align}
where the sum runs over all pairings of spins $i$ and $j$ that are normal to, but not lying on, the ring.
There are four symmetry-inequivalent pairings, here labeled $1,2A,2B,3$, corresponding to nearest, second-nearest and third-nearest neighbour normal spins.

Note that this expression could have been equivalently written in terms of the radial spins (green dots in Fig.~\ref{fig:normalspins}b). The equivalence stems from the fact that we work in the 2I2O manifold: 
if $S^z_n$ and $S^z_r$ are, respectively, normal and radial spins belonging to a tetrahedron consisting of spins $n,r,1,2$, the 2I2O constraint may then be used to replace $S^z_r$ by $-S^z_n-S^z_1-S^z_2$. If sites 1 and 2 coincide with a ring-flip, the factors of $S^z$ may be replaced by eigenvalues as described in Sec.~\ref{sec:perturb}.
%
%
%

\subsection{Relevance of $H_{\pm\pm}$}
\label{app:hpmpm}
\begin{figure}
\includegraphics{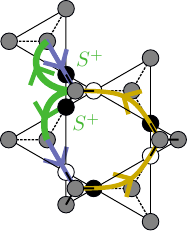}
\caption{An example of leading-order contribution of $H_{\pm\pm}$ to complex ring flip processes.}
\label{fig:hpmpm}
\end{figure}

In principle, the spinon sublattice mixing from $H_{\pm\pm}$ may also contribute to $g'$. However, as we mentioned in the main text, such contributions are higher order than those from $J_{z\pm}$ and have been neglected in our work. Effective operators within the ice ensemble must annihilate all virtual spinons they create. $H_{\pm\pm}$ makes this task particularly difficult: It creates doubly-charged spinons, necessitating additional terms to return to the spin ice ensemble. The lowest-order term capable of annihilating these spinons by nontrivial loop is of the form $H_{z\pm}^2H_{\pm\pm}H_{\pm}^2$, sketched in Fig.~\ref{fig:hpmpm}. All such terms are higher order in $1/J_{zz}$ than the ring flips considered in the main text, so they may be justifiably dropped.
%
%
%









%
%

\section{Chart dependence and vison definition}
\label{app:chart_indep}
In Sec.~\ref{sec:periodicity}, we wrote down a local definition of a vison in terms of the winding number of four phases. It should be stressed that this notion of a vison is chart dependent. For example, consider an $A$-sublattice void bounded by the four plaquettes $p_1, p_2, p_3, p_4$ with corresponding Wilson loops $e^{ie_{p_4}} =\exp[i\,(\pi/2 +3\epsilon)]$, $e^{ie_{p_1}}=e^{ie_{p_2}}=e^{ie_{p_3}}=e^{i\,(\pi/2-\epsilon)}$, $0<\epsilon\ll1$. This structure is considered to have a vison in the chart $e \in (-\pi,\pi)$, but has no vison in the chart $e \in (\pi/2,-3\pi/2)$. This presents an issue if we wish to treat these vortex states as `matter', decoupled from the gauge field -- it seems that we can fluidly move between them by a trivial change of coordinates. 

There is no such problem in a continuum gauge field. Starting from the closed surface bounding a void $v$ of the diamond lattice, we can arbitrarily subdivide each plaquette into infinitesimally small sub-plaquettes, ultimately converging to the integral
$$
\frac{1}{2\pi} \int_{\partial v} \nabla\times \boldsymbol{a} \cdot d\boldsymbol{S}
\, , 
$$
which is Dirac quantized in the usual way. In the winding-number language, we subdivide our straight line phases into arbitrarily small segments, ultimately reaching a point where each small plaquette carries an arbitrarily small field.
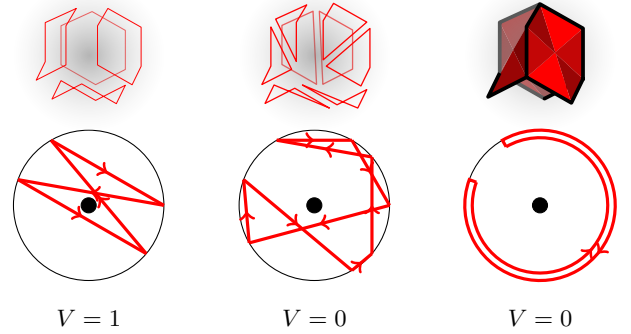
\begin{figure}
\begin{tikzpicture}
    \draw (0,0) circle (1) ++ (0,-1.5)node{$V=1$} ++(0,3.5) node{\includegraphics[width=2cm]{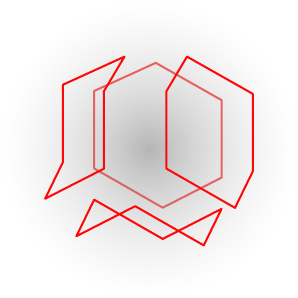}};
\filldraw (0,0) circle(0.1);
\begin{scope}[very thick,red, xshift=0cm, decoration={
    markings,
    mark=at position 0.5 with {\arrow{>}}}
    ] 
    \draw[postaction={decorate}] (0:1)--(160:1);
    \draw[postaction={decorate}] (160:1) -- (320:1);
    \draw[postaction={decorate}] (320:1) -- (120:1);
    \draw[postaction={decorate}] (120:1) -- (0:1);
\end{scope}
\draw (3,0) circle (1) ++ (0,-1.5)node{$V=0$} ++(0,3.5) node{\includegraphics[width=2cm]{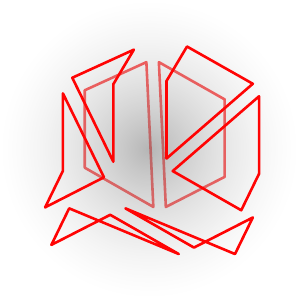}};;
\filldraw (3,0) circle(0.1);
\begin{scope}[very thick,red, xshift=3cm, decoration={
    markings,
    mark=at position 0.5 with {\arrow{>}}}
    ] 
    \draw[postaction={decorate}] (0:1)--(210:1);
    \draw[postaction={decorate}] (210:1)--(160:1);
    \draw[postaction={decorate}] (160:1) -- (300:1);
    \draw[postaction={decorate}] (300:1) -- (320:1);
    \draw[postaction={decorate}] (320:1) -- (40:1);
    \draw[postaction={decorate}] (40:1) -- (120:1);
    \draw[postaction={decorate}] (120:1) -- (60:1);
    \draw[postaction={decorate}] (60:1) -- (0:1);
\end{scope}
\draw (6,0) circle (1) ++ (0,-1.5)node{$V=0$} ++(0,3.5) node{\includegraphics[width=2cm]{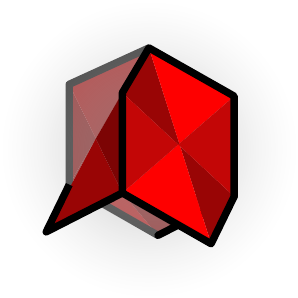}};;
\filldraw (6,0) circle(0.1);
\begin{scope}[very thick,red, xshift=6cm, decoration={
    markings,
    mark=at position 0.5 with {\arrow{>}}}
    ] 
    \draw[postaction={decorate}] (120:0.9)--(120:1) arc (120:-200:1) -- (-200:0.9);
    \draw[postaction={decorate}] (-200:0.9) arc (-200:120:0.9);
\end{scope}
\end{tikzpicture}
\caption{Subdivision of a void-shaped cell in a continuum $U(1)$ gauge theory on length scales below lattice spacing.}
\end{figure}

A partial resolution of this chart dependence on the lattice comes from carefully constructing a coarse-grained continuum limit.
Suppose for now that the system contains only zero fluxes. Let us then introduce a defect on some void $v$ using a half-infinite Dirac string, and fix some $\epsilon>0$ to be a small parameter. There exists a large volume $\Omega$ containing the point defect, such that all elementary plaquettes making up its surface $\partial \Omega$ satisfy $|e| < \epsilon$, which may be verified by noting that at large separations $e \sim 1/r^2$. This procedure resolves the ambiguity associated with assigning directions to the four phases of Fig.~\ref{fig:vison_sketch}: Any large phases are simply subdivided by taking larger Gaussian surfaces until all fields are small. In the limit $\epsilon \to 0^+$, the vison charge
approaches an integrated field curvature:
\begin{align*}
    \sum_{v\in \Omega} \nabla\cdot_v e &= \sum_{p \in \partial \Omega} e_p n_p \ \to\ \int_\mathbf{\partial\Omega} \boldsymbol{e} \cdot d\boldsymbol{S}
\end{align*}
We have defined $n_p = \pm 1$ to be orientation consistent with the outward normal of the volume $\Omega$. We recognize this integral to be the usual Chern number~\cite{luscherTopologyAxialAnomaly1999} of the Dirac monopole.
The argument presented here does not generally apply in the presence of multiple visons, which ultimately place a lower bound on $\epsilon$ depending on their separation. Indeed, the visons become increasingly ill-defined as their mean separation approaches the lattice scale.
%
%

\section{Chart-independent static structure factors}
\label{app:ssf_chart_indep}
The $\mathbb{Z}_2$ duality transform $\mathcal{R}_{a_{\pi/2}} \mathcal{C}$ interchanges the phases $U(1)_0 \leftrightarrow U(1)_{\pi/2}$, suggesting that the physics of the model should be the same for $\phi_g=\theta$ and $\phi_g = \pi/2 - \theta$. It is however not immediate to see this symmetry in the static electric-sector correlators $\langle e e \rangle$, illustrated in Fig.~\ref{fig:ssf}, as a consequence of having chosen a specific chart, namely the $(-\pi,\pi)$ one (see Sec.~\ref{sec:periodicity}). 

\begin{figure*}
    \includegraphics{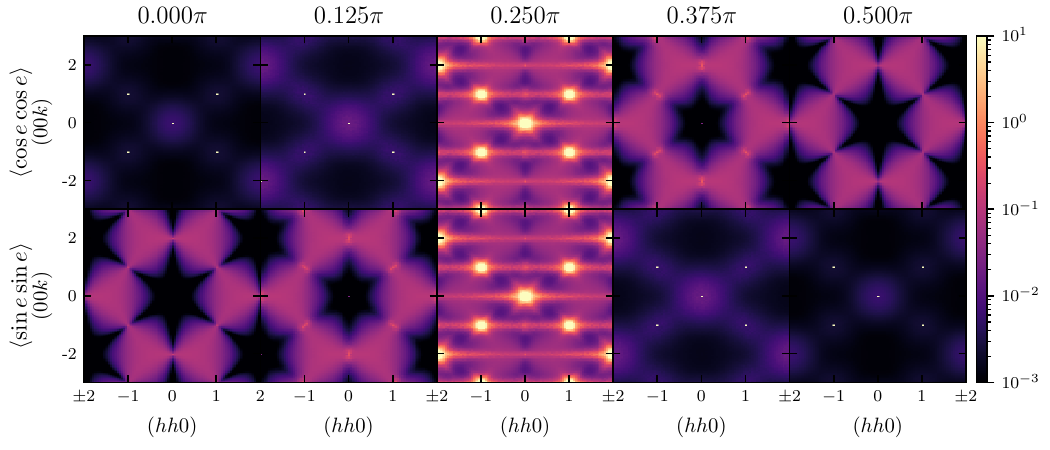}
    \caption{Evolution of static electric field correlations from Fig.~\ref{fig:ssf} in terms of chart-independent Wilson loop quantities. Plot titles reflect the value of $\phi_g$.
    Temperature $T/\rho_g=0.5$ is held constant.
    Bragg peaks are present at (000) in all plots, excepting $(\phi_g=0, \langle \cos e \cos e\rangle)$ and $(\phi_g=\pi/2, \langle \sin e \sin e\rangle)$; they may, however, be difficult to discern due to rasterizaton.}
    \label{fig:ssf_cos_sin}
\end{figure*}

Fig.~\ref{fig:ssf_cos_sin} shows the static structure factors as seen through manifestly chart invariant plaquette correlations, $\langle \sin e \sin e \rangle$ and $\langle \cos e \cos e \rangle$. Where $\phi_g < \pi/2$, we see $\langle \sin e \sin e \rangle$ correlations with very similar features to $\langle e e \rangle$ (Fig.~\ref{fig:ssf}). Similarly, when $\phi_g>\pi/2$, we see that $\langle \cos e \cos e \rangle \simeq \langle (e + \pi/2)(e + \pi/2) \rangle$ reveals (shifted) $\langle e e \rangle$ correlations.  
At $\phi_g =0, 0.125\pi$, the $\langle \cos e \cos e \rangle$ correlations reveal the Bragg peaks at (000) and (111) expected of a static order on the pyrochlore lattice. Both the (000) and (111) peaks are surrounded by Gaussian diffuse features corresponding to thermal fluctuations. Though the energy scale $\rho_g = \sqrt{g^2 + (g')^2}$ is held constant in this plot, the strength of the fluctuations is controlled by $T/g$ rather than $T/\rho_g$, and the diffuse features are therefore more intense at $\phi_g = 0.125$. Of course, an identical discussion applies to the $\sin e$ correlations at $\phi_g = 0.375\pi, 0.5\pi$. 

\begin{figure}
    \includegraphics[]{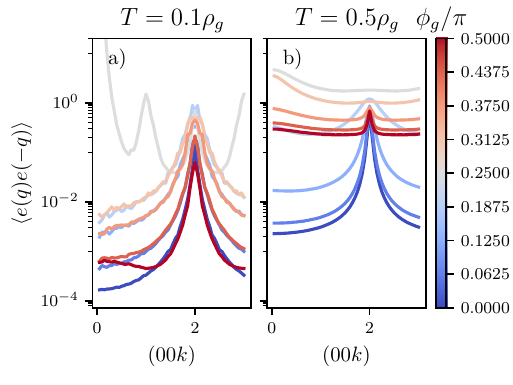}
    \caption{Evolution of electric field correlator $\langle e(\boldsymbol{q})e(-\boldsymbol{q}) \rangle$ along the [00k] line with varying $\phi_g$. Panel a) shows the crossing of the sharp phase transition; panel b) a smooth crossover above the critical temperature. Symmetry guarantees no photon contribution along this cut. The Bragg peak at $k=0$ was explicitly cut off from this plot. The simulation parameters between panels a) and b) are identical; the low-temperature system has domain walls that cause oscillations in $k$ space.}
    \label{fig:ssf_cut}
\end{figure}

Yet another view of the phase transition comes from Fig.~\ref{fig:ssf_cut}, showing cuts along the line $h=0$. Photon correlations are known to vanish exactly along this line; this observable probes only effects due to the visons. The Debye screening of the photons manifests as a Lorentzian centred on (002), of characteristic width set by the Debye length (itself a function of vison density)~\cite{szaboSeeingLightVison2019}. As one may expect, the vison density is maximal at the phase transition where the field tension $\mathcal{T} = \max\{|g|,|g'|\}$ is smallest relative to the temperature. Thermally excited, isolated vison dipoles manifest as a uniform background in $k$-space: indeed, this background is largest at the degeneracy point $\phi_g=\pi/4$.

At low temperature (Fig.~\ref{fig:ssf_cut}a) we see that traces at $\phi_g$ and $\pi/2 - \phi_g$ overlap for most of the Brillouin zone, differing only by the presence of a Gaussian diffuse peak at (000). This feature is due to fluctuations in the Bragg peak at (000) in the $U(1)_{\pi/2}$ phase. Note that the duality transform $\mathcal{R}_{a_{\pi/2}}\mathcal{C}$ maps $e\mapsto \pi/2 - e$, preserving all non-singular correlations.
At higher temperature (see panel~\ref{fig:ssf_cut}b), the chart boundary for $e$ distorts this picture: when a fluctuation in $e$ is large enough to cross the chart boundary at $\pi$, the fluctuation manifests in the correlation function as a uniform background, associated with a nearest-neighbour vison dipole~\cite{szaboSeeingLightVison2019}. This issue is more pronounced in the $U(1)_\pi/2$ phase than the $U(1)_0$ one due to the closer proximity of the chart boundary to the mean value of $e$.
%
%
%
%
%

\section{Semiclassical simulation}
\label{app:numerics}
Our Monte Carlo algorithm is identical to that established by Szab\'{o} and Castelnovo~\cite{szaboSeeingLightVison2019}.
The method is based on classical Monte Carlo simulations of the large-$S$ path integral formulation of the $U(1)$ gauge theory, implemented by a Metropolis algorithm consisting of the following stages: 
\begin{enumerate}
    \item[MC1] Apply a random gauge transformation to every tetrahedron.
    \item[MC2] Attempt to rotate spins about the local $z$ axis, accepting the move with probability $\propto \exp(-\beta \Delta E)$
    \item[MC3] `Ring flip' step: attempt to change the $z$ components of the six spins around a plaquette by some angle $\delta$ in an alternating ring flip pattern. For a hexagon $p$ of spins labelled by $1,2,3,4,5,6$, this is the transformation
    $$
    \left\{
    \begin{array}{ccc}
            S^z_1  \mapsto S^z_1 + \delta, &
            S^z_2  \mapsto S^z_2 - \delta,&
            S^z_3  \mapsto S^z_3 + \delta,\\
            S^z_4  \mapsto S^z_4 - \delta,&
            S^z_5  \mapsto S^z_5 + \delta,&
            S^z_6  \mapsto S^z_6 - \delta
    \end{array} \right\} \, .
    $$  
    The move is either rejected if any $S^z$ is moved outside of $[-1, 1]$, or accepted with probability $\propto \exp(-\beta \Delta E)$. This step mimics the infinitesimal, large-$S$ action of the $\mathcal{O}_p$ operator.
\end{enumerate}
Depending on the observable we are attempting to measure, we employ the following two basic processes. 

\paragraph{Annealing.}
Starting from $1/\beta > \rho_g$, we run a fixed number $N_{\rm step}$ of Metropolis steps and then lower the temperature. We use $N_{\rm temp}$ temperatures, which are logarithmically spaced between $T_{\rm hot}$ and $T_{\rm cold}$. These two variables must be tuned carefully. To simulate a realistic annealing process, one should choose $N_{\rm step}$ to be large enough for the system to explore the full configuration space. Moreover, if $N_{\rm step}$ is chosen to be exceedingly small, one effectively quenches the system into a metastable state. Such metastability is unavoidable in the critical regime, where costly domain walls are kinetically prevented from annihilating (cf. Fig.~\ref{fig:ssf} and Sec.~\ref{sec:ssf}). 

\paragraph{Time evolution.}
The semiclassical equation of motion  
\begin{align}
    \partial_t\bm{S}_i = \frac{\partial H_{\text{eff}}[\{\bm{S}\}]}{\partial \bm{S}_i} \times \bm{S}_i
    \label{eq:time_ev}
\end{align}
is integrated numerically while time series for relevant correlators are stored. As an order of magnitude estimate for the time taken by Monte Carlo to explore the full phase space, we exploit the emergent speed of light $c_{\rm{QSI, large-}S} \sim 0.15 g a_0$~\cite{kwasigrochSemiclassicalApproachQuantum2017}: for a photon to complete a full loop on
the three-torus of side $8La_0$ (when $g'=0$), one requires $\rho_g t = \frac{8L}{0.15}$, i.e., we must capture at
least $\sim 50 L$ oscillation periods.

Time-evolved world lines of the system are then averaged over the thermal ensemble of initial states, essentially computing a Monte Carlo summation of a zeroth-order approximation to the real-time path integral.
%
%

\section{Electric - magnetic conventions}
\label{sec:EM_duality}
The classical Maxwell equations in the presence of magnetic monopoles read (in Gauss units)
\begin{align}\label{eq:maxwell1}
    \bm{\nabla} \cdot \bm{e} &= 4\pi \rho_e\\
    \bm{\nabla} \cdot \bm{b} &= 4\pi \rho_m\\
    \bm{\nabla} \times \bm{e} + \frac{1}{c}\frac{\partial\bm{b}}{\partial t} &= -\frac{4\pi}{c} \bm{J}_m\\
    \bm{\nabla} \times \bm{b} - \frac{1}{c}\frac{\partial \bm{e}}{\partial t} &= \frac{4\pi}{c}\bm{J}_e
\, . 
\end{align}
These equations constrain the transformation properties of $\bm{e}$ and $\bm{b}$ under discrete symmetries. The modified Faraday law implies that $\partial_t \bm{b}$, $\nabla \times \bm{e}$, and $\bm{J}_m$ must all transform the same way under parity and time reversal. This in turn guarantees that $\bm{e}$ and $\bm{b}$ have opposite properties under these transformations, i.e., precisely one of the two is a vector and precisely one is even under time reversal.

In conventional electromagnetism, we take $\bm{e}$ to be a $\mathcal{T}$-even vector, automatically making $\bm{b}$ a TRS-odd pseudovector. The spin ice literature is divided on this assignment - the convention established by Hermele et al.~\cite{hermelePyrochlorePhotonsSpin2004} calls the coarse-grained $S^z$ moments, a TRS odd vector, an electric field; the convention established by Castelnovo et al.~\cite{castelnovoMagneticMonopolesSpin2008} calls this the magnetic field. Neither convention is ultimately equivalent to Maxwell theory.
\begin{table}
\begin{tabular}{l|c|c|c}
    &
    Hermele et al.~\cite{hermelePyrochlorePhotonsSpin2004}    &
    Castelnovo et al.~\cite{castelnovoMagneticMonopolesSpin2008} & 
    Maxwell~\cite{maxwellVIIIDynamicalTheory1865}
   \\
\hline 
    $\bm{e}$ & $S^z_i \bm{n}_{\mu(i)}$ & $\left(\nabla_p \times \phi \right) \bm{n}_{\mu(p)}$ 
    \\
    \hline    
    $\bm{b}$ & $\left(\nabla_p \times a \right)\bm{n}_{\mu(p)}$ & $S^z_i \bm{n}_{\mu(i)}$
    \\
    \hline                    
    $\mathcal{T}\bm{e}(\bm{r}, t)$ &
        $-\bm{e}(\bm{r}, -t)$ &
        $+\bm{e}(\bm{r}, -t)$ &
        $+\bm{e}(\bm{r}, -t)$ 
    \\
    $\mathcal{T}\bm{b}(\bm{r}, t)$ &
        $+\bm{b}(\bm{r}, -t)$ &
        $-\bm{b}(\bm{r}, -t)$ &
        $-\bm{b}(\bm{r}, -t)$ 
    \\
    \hline
    $\mathcal{I}\bm{e}(\bm{r}, t)$ &
        $+\bm{e}(-\bm{r}, t)$ &
        $-\bm{e}(-\bm{r}, t)$ &
        $+\bm{e}(-\bm{r}, t)$ 
    \\
    $\mathcal{I}\bm{b}(\bm{r}, t)$ &
        $-\bm{b}(-\bm{r}, t)$ &
        $+\bm{b}(-\bm{r}, t)$ &
        $-\bm{b}(-\bm{r}, t)$ 
    \\
\hline                                                      
\end{tabular}
\caption{Symmetry conventions.}
\end{table}       

The units of (emergent) electomagnetism we work in are Gaussian. Explicitly, the units of electric charge $q_e$ and magnetic charge $q_m$ are defined such that the vison-vison and spinon-spinon interaction energies are 
\begin{align}
    U_m(R) &= \frac{q_m^2}{R}\\
    U_e(R) &= \frac{q_e^2}{R}
\, , 
\end{align}
as the quasiparticle separation $R\to\infty$. 
%
%

\section{Large-$S$ perturbation theory in $g'/g$
\label{app:continuum_calculation}} Starting from a system in the $U(1)_0$ phase, we develop a quantitative understanding of the perturbative effect of $g'$ on the system. We will show that the $\sin e$ term acts as an anharmonic medium for the photons analogous to a nonlinear crystal. 
The Hamiltonian of Eq.~\eqref{eq:large_S_ham} is first expanded about the 0-flux vacuum in falling powers of $1/\tilde{s} = (S+1/2)^{-1}$, where it is noted that 
$e \sim S^z/\tilde{s} \sim \tilde{s}^{-1/2}$, and then separated into quadratic and interacting parts:
\begin{align}
H_0[a] &= \frac{gz}{2} \sum_j\left(\frac{S^z_j}{\tilde{s}}\right)^2 + \sum_p \frac{g}{2}e_p^2 + \sum_v g' 2\pi V(v)\ \label{eq:quadratic},
\\
H_I[a] &= \sum_p -\frac{g'}{3!}e_p^3 - \frac{g}{4!}e_p^4 + \frac{g}{6!}e_p^6 + \ldots \, . 
\end{align}
By taking the expansion, we are supposing that $e_p$ is in some sense small, and it is therefore safe to assume in this section that $e_p$ and $\nabla\times_p a$ are essentially synonymous. (Strictly speaking, a chart is being chosen for $a$ in addition to $e_p$.) 
As noted in the main text (Sec.~\ref{ssec:largeS-main}), $V(v)$ in the large-$S$ limit is a non-dynamical topological term decoupled from the gauge-field dynamics, and it may be set to zero for the purposes of this calculation.
The term proportional to $gz$ in Eq.~\eqref{eq:quadratic}, where $z=6$ counts the number of spins on a plaquette, arises from the Villain expansion of the ring-flip~\cite{hermelePyrochlorePhotonsSpin2004,kwasigrochSemiclassicalApproachQuantum2017,kwasigrochVisongeneratedPhotonMass2020}.
\\
It can be shown that the next-leading, mixed-operator corrections from $g'$ are of the form $g' (S^z/\tilde{s})^2 e_p$. In the spirit of presenting a minimal effective theory that captures the high-energy features from semiclassical numerics, such terms have been dropped from $H_I$.
These terms essentially amount to a renormalization of the three-photon scattering vertex $e^3$ already discussed in the main text (Sec.~\ref{ssec:largeS-main}).
%

The partition function of the system can finally be written as: 
\begin{align}
\mathcal{Z} &= \int\mathcal{D}a\ \exp(-\mathcal{S}_0[a] -\mathcal{S}_I[a]) \\
\mathcal{S}_0[a] &= \int d\tau\  H_0[a]\ , \label{eq:quadratic_action}\\
\mathcal{S}_I[a] &= \int d\tau\  H_I[a] \, . 
\end{align}
We compute the imaginary-time correlator $\langle S^z(q, \tau) S^z(-q, 0)\rangle $ by expanding the interaction $\exp({-\mathcal{S}_I[a]}) =1 + \sum_{k=1}^\infty \mathcal{S}_I^{(k)}[a]$ where $\mathcal{S}_I^{(k)}$ contains only $k$-th powers of $a$:
\begin{widetext}
\begin{align}
    \langle S^z(\mathbf{q}, \tau)S^z(-\mathbf{q}, 0) \rangle &=  \frac{1}{\mathcal{Z}}{\int \mathcal{D}a\: S^z(\mathbf{q},\tau)S^z(-\mathbf{q},0) 
    \left[1 + \mathcal{S}_I^{(4)}[a] + \mathcal{S}_I^{(6)}[a]\right] e^{-\mathcal{S}_0[a]}}
    \\
    \mathcal{S}_I^{(4)}[a] &= \int d\tau\ \sum_p -\frac{g}{4!}e_p(\tau)^4
    \\
    \mathcal{S}_I^{(6)}[a] &= - \int d\tau\ \frac{g}{6!}\sum_p e_p(\tau)^6 + \frac{1}{2}\left(\frac{g'}{3!}\right)^2\int d\tau_1d\tau_2 \sum_{p_1,p_2}e_{p_1}(\tau_1)^3 e_{p_2}(\tau_2)^3
\, . 
\end{align}
\end{widetext}
Note in particular that, due to the form of $H_I$, $\mathcal{S}_I^{(k)}$ contains only powers of $a^3$ and higher. Due to the vanishing of odd-powered terms in the action, the leading order in the series is then $\mathcal{S}_I^{(4)}$. 
The Hartree-Fock corrections arising from $\mathcal{S}_I^{(4)}[a]$ were calculated in Ref.~\onlinecite{kwasigrochSemiclassicalApproachQuantum2017}. They correct the energy of the existing photon modes and renormalise the speed of light, but do not introduce any additional modes. We shall therefore neglect them hereafter. 

It is necessary at this point to move from indexing by links and plaquettes to indexing by tetrahedron and pyrochlore sublattice. We index a spin uniquely by the position $\mathbf{R}_t$ of its nearest $A$-type tetrahedron and its pyrochlore sublattice $\lambda \in \{0,1,2,3\}$. Similarly, a plaquette may be uniquely indexed by its nearest-neighbour $A$-void location $\mathbf{R}_v$ and its plaquette sublattice $\mu \in\{0,1,2,3\}$. It may be readily verified that the spin $(\mathbf{R}_t, \lambda)$ is located at the position $\mathbf{R}_t + (1-\chi) \mathbf{b}_\lambda /2$, and the centre of the plaquette $(\mathbf{R}_v, \mu)$ lies at $\mathbf{R}_v - (1-\chi/3) \mathbf{b}_\mu /2$, where $\chi \in [0,1)$ is the breathing anisotropy of the original pyrochlore lattice. For notational simplicity we will set $\chi=0$, as it has only a marginal effect on the derivation to follow.

Following the standard QED approach~\cite{kwasigrochSemiclassicalApproachQuantum2017, bentonSeeingLightExperimental2012}, we introduce bosonic operators associated with each site $c_{\mathbf{R}_t, \lambda}$, $[c_{\mathbf{R}_t, \lambda}, c^\dagger_{\mathbf{R}_s, \rho} ] = \delta_{ts} \delta_{\lambda\rho}$ and define their associated Fourier modes to be $c_{\lambda}(\mathbf{k})= \sqrt{2/N_t}\sum_{t\in A} \exp(-i\mathbf{k}\cdot\mathbf{R}_t) c_{\mathbf{R}_t, \lambda}$. Here $N_t$ is the total number of tetrahedra, which is equal to the total number of voids. Inspired by QED, we make the definitions
%
%
\begin{align}
a_{\lambda}(\mathbf{k},t) &= 
\frac{1}{\sqrt{2 \tilde{s} \epsilon_{\mathbf{k}}}}
\left[
  c_{\lambda}^{\dagger}(\mathbf{k},t) 
  + c_{\lambda}(-\mathbf{k},t)
\right] 
\\
\Pi_{\lambda}(\mathbf{k},t)&= 
-\frac{\partial a_\lambda(\mathbf{k})}{\partial t} 
= i\sqrt{\frac{\epsilon_{\mathbf{k}}}{2\tilde{s}}}
\left[
  c_{\lambda}^{\dagger}(\mathbf{k},t) 
  - c_{\lambda}(-\mathbf{k},t)
\right]
\, , 
\end{align}
where $\Pi = S^z/\tilde{s}$ is now understood as the canonical momentum conjugate to $a$, and $\epsilon_{\mathbf{k}}$ is to be determined. The Fourier transforms of $e$ and $\Pi$ may then be written in terms of the $a$ variables,
\begin{widetext}
\begin{align*}
    e\left(\mathbf{R}_v - \frac{\mathbf{b}_\mu}{2}\right) &= \sum_{\eta = \pm 1}\sum_{\lambda \neq \mu} \eta (-)^{\lambda - \mu} a\left(\mathbf{R}_v- \frac{\mathbf{b}_\mu}{2} + \eta \frac{a_0 \mathbf{b}_\mu \times \mathbf{b}_\lambda }{\sqrt{8} \|\mathbf{b}_\mu \times \mathbf{b}_\lambda\|}\right) 
    =
    \sqrt{\frac{2}{N_{t}}}\sum\limits_{\mathbf{k}\in \text{BZ}, \lambda}Z_{\mu\lambda}(\mathbf{k})e^{i \mathbf{k} \cdot (\mathbf{R}_v-\mathbf{b}_{\mu}/2)}a_{\lambda}(\mathbf{k}) \\
    \Pi(\mathbf{R}_t + \mathbf{b}_\lambda/2) &= \sqrt{\frac{2}{N_t}}\sum\limits_{\mathbf{k}\in\text{BZ}} \Pi_{\lambda}(\mathbf{k})e^{i \mathbf{k}\cdot(\mathbf{R}_{t}+ \mathbf{b}_{\lambda}/2)}
\, , 
\end{align*}
where 
\begin{align*}
    Z_{\mu\lambda}(\mathbf{k}) &= -2i \sin\left( \frac{a_0}{ \sqrt{8} } \frac{\mathbf{k} \cdot \mathbf{b}_\mu \times \mathbf{b}_\lambda }{ | \mathbf{b}_\mu \times \mathbf{b}_\lambda |} \right) 
\, . 
\end{align*}
%
After making this substitution, the Wick rotated quadratic action Eq.~\eqref{eq:quadratic_action} reads
%
\begin{align}
    S_0[a] &= \int \frac{dt}{2\tilde{s} \epsilon_{\mathbf{k}}} \sum_{\mathbf{k}} \left\{ \left[\epsilon_{\mathbf{k}}^2 + \frac{1}{z}  Z(\mathbf{k})^2\right]_{\lambda\lambda'} 2c^\dagger_{\lambda}(\mathbf{k}) c_{\lambda'}(\mathbf{k}) 
+
\left(\epsilon_{\mathbf{k}}^2 - \frac{1}{z}  Z(\mathbf{k})^2\right)_{\lambda\lambda'} \left[ c_{\lambda}(\mathbf{k}) c_{\lambda'}(-\mathbf{k}) + c^\dagger_{\lambda}(\mathbf{k}) c^\dagger_{\lambda'}(-\mathbf{k}) \right] 
    \right\}
\, ,
\end{align}
\end{widetext}
from which we see that the anomalous terms may be canceled by diagonalizing $z^{-1/2} Z$ and setting $\epsilon_{\mathbf{k}}$ to its unique nonvanishing (doubly degnerate) eigenvalue. This fixes $\epsilon_{\mathbf{k}}$, recovering the standard photon dispersion. 
%
The novel term introduced by the $g'$ coupling arises from the third order term in the expansion of $\sin e$,
\begin{align}
V_{\rm{eff}} &= \frac{g'}{3!}\sum\limits_{\mathbf{R}_v,\mu} e\left(\mathbf{R}_v-\mathbf{b}_\mu/2\right)^{3} \nonumber \\
 &= \frac{g'}{3!} \sum_{\mathbf{k}_1,\mathbf{k}_2, \mathbf{k}_3} \Gamma^{\mu_1\,\mu_2\,\mu_3}_{\mathbf{k}_1\, \mathbf{k}_2\, \mathbf{k}_3} a_{\mu_1}(\mathbf{k}_1) a_{\mu_2}(\mathbf{k}_2) a_{\mu_3}(\mathbf{k}_3)
\, ,
\end{align}
where
\begin{align}
\Gamma^{\mu_1\,\mu_2\,\mu_3}_{\mathbf{k}_1\, \mathbf{k}_2\, \mathbf{k}_3} 
&=\frac{\delta_{\mathbf{k}_1+ \mathbf{k}_2+ \mathbf{k}_3}}{\sqrt{N_t/2}}
\sum_\lambda Z_{\lambda\mu_1}(\mathbf{k}_1)Z_{\lambda\mu_2}(\mathbf{k}_2)Z_{\lambda\mu_3}(\mathbf{k}_3)
\, . 
\label{eq:Gamma_def}
\end{align}

At order $S^{(6)}$, there are two $g'$-mediated diagrams in the photon self energy (Fig.~\ref{fig:feyn_diag}). The bubble diagram has no contribution, as the propagator connecting the two vertices carries zero momentum, and from the form of $\Gamma$ in Eq.~\eqref{eq:Gamma_def} this contribution must vanish (recall that $Z(\mathbf{0})$ is the zero matrix).
The remaining diagram gives a contribution to the spectral weight that, after summing over Matsubara frequencies, reduces to
\begin{widetext}
\begin{align}
\langle S^z_\sigma(\mathbf{q},\omega) S^z_{\tau}(-\mathbf{q},0)\rangle &= \tilde{s}^2\langle \Pi_\sigma(\mathbf{q},\omega) \Pi_\tau(-\mathbf{q},0) \rangle 
\nonumber \\
&= \frac{\epsilon_{\mathbf{p}}\epsilon_{\mathbf{q}-\mathbf{p}}}{(2\tilde{s})^2} \frac{1}{2}  \left(\frac{g'}{3!}\right)^{2}\sum\limits_{\mathbf{p}\in \text{BZ}}\sum_{\lambda,\xi=0}^3 \delta(\epsilon_{\mathbf{p}} + \epsilon_{\mathbf{q}-\mathbf{p}} - i \omega) 
\Gamma\begin{smallmatrix}\sigma & \lambda & \xi\\
                -\mathbf{q}& \mathbf{q}-\mathbf{p}& \mathbf{p}\end{smallmatrix}
    \Gamma\begin{smallmatrix}\tau & \lambda & \xi \\ \mathbf{q}& \mathbf{p}-\mathbf{q}& -\mathbf{p}\end{smallmatrix}
    \left[ \vphantom{\sum}
      1 + n_B(\epsilon_\mathbf{p}) + n_B(\epsilon_{\mathbf{q}-\mathbf{p}})
    \right]
\, . 
\label{eq:analytic_expresion_vertex}
\end{align}
\end{widetext}
Hidden in the structure of the $\Gamma$ terms is a symmetry-enforced Kronecker delta between $\mu$ and $\rho$.
We plot (the trace of) this this term in Fig.~\ref{fig:upconversion}, with an arbitrary intensity scale.

\end{document}